\documentclass[12pt,preprint]{aastex}
\begin{document}
\title{Estimates of Stellar Weak Interaction Rates for Nuclei in the Mass Range A=65-80}
\author{Jason Pruet}
\affil{Lawrence Livermore National Laboratory, L-414, P.O. Box 808, 
Livermore, CA 94551}
\author{George M. Fuller}
\affil{Department of Physics, University of California, 
San Diego, La Jolla, CA 92093-0319}

\begin{abstract}
We estimate lepton capture and emission rates, as well as
neutrino energy loss rates, for nuclei in the mass range
A=65-80. These rates are calculated on a temperature/density grid
appropriate for a wide range of astrophysical applications including
simulations of late time stellar evolution and x-ray bursts.  The
basic inputs in our single particle and empirically inspired model 
are i)experimentally measured level information, weak
transition matrix elements, and lifetimes, ii) estimates of matrix
elements for allowed experimentally-unmeasured transitions based on
the systematics of experimentally observed allowed transitions, and
iii) estimates of the centroids of the GT resonances motivated by
shell model calculations in the fp shell as well as by (n,p) and (p,n)
experiments. Fermi resonances (isobaric analog states) are also included, 
and it is shown
that Fermi transitions dominate the rates for most interesting proton
rich nuclei for which an experimentally-determined ground state lifetime is
unavailable. For the purposes of comparing our results with more
detailed shell model based calculations we also calculate weak rates
for nuclei in the mass range A=60-65 for which Langanke and
Martinez-Pinedo have provided rates. The typical deviation in the
electron capture and beta decay rates for these $\approx 30$ nuclei is
less than a factor of two or three for a wide range of temperature and
density appropriate for pre-supernova stellar evolution. We also
discuss some subtleties associated with the partition functions used
in calculations of stellar weak rates 
and show that the proper treatment of the
partition functions is essential for estimating high temperature beta
decay rates. In particular, we show that partition functions based on 
un-converged Lanczos calculations can result in estimates of high temperature 
beta decay rates that are systematically low.
\end{abstract}

\section{Introduction}

In this paper we provide estimates for weak interaction rates
involving intermediate mass nuclei. \cite{aufder1} and \cite{aufder2}
have argued that at late times the electron fraction in the Fe core of
pre-supernova stars can be so low that weak processes involving the
A$>65$ nuclei we study are important.  Depending on the entropy per
baryon, which determines the free proton fraction, electron capture on
heavy nuclei may also play an important role during collapse
\citep{bethe,fullerapj}. In addition, the weak rates we provide for proton
rich nuclei may be used in studies of nucleosynthesis and energy
generation in x-ray bursts and other {\it rp-}process sites \citep{woosrp}.
For the {\it rp-}process, weak rates are needed for proton rich nuclei at
least up to mass 110. Electron capture and positron decay rates for
proton rich nuclei in the mass range A=81-110, as well as a discussion
of some peculiarities of the weak rates in the {\it rp-}process environment,
will be presented in another paper (Pruet and Fuller, in preparation).

The formidable task of calculating the electron and positron capture
and emission rates in the conditions characteristic of these
astrophysical environments has received more than four decades of
attention. The first self consistent calculations to include
the effects of the Fermi and Gamow-Teller (GT) resonances as well as the
thermal population of these resonances for a broad
range of nuclei and thermodynamic conditons were done by
\cite{FFN1,FFN2,FFN3} (hereafter FFN), 
\cite{fullerapj}, and \cite{fuller4}.  FFN presented a
physically intuitive and computationally tractable method for
determining the strength and excitation energy of the Fermi and GT
resonances. The groundwork for the treatment of these resonances in a
thermal environment was laid down by \cite{bethe} and FFNII.  The
importance of first forbidden transitions, blocking, and thermal
unblocking for the very neutron rich nuclei present during collapse
was pointed out by \cite{fullerapj}.
\cite{cooperstein} calculated electron capture rates for these nuclei
by estimating the parity forbidden matrix elements as well as the 
effects of thermal unblocking of the allowed ${\rm GT^+}$ strength. 

The last two decades have seen a great increase in our understanding
of weak interaction systematics in intermediate mass nuclei. There are now
semi-direct measurements of the GT strength distribution for roughly
40 nuclei from forward angle (n,p) and (p,n) scattering
experiments. There are also shell model calculations for the strength
distribution for some 400 nuclei in the fp shell \citep{lan1,lan2}, as
well as a number of RPA and QRPA calculations of the GT resonance in
heavier nuclei.  One consequence of these studies is that the FFN
prescription misses some important systematics in the assigning the
centroid of the GT resonance.  The influence of this misassignment on
the rates was discussed by \cite{mathew1}, \cite{lan1}, and
\cite{lan2} (hereafter LMP). \cite{lan2} also provided updated calculations of the
rates for A$\leq$ 65 based on large scale shell model calculations.

We attempt to remedy the misassignment of the GT centroids. Other than this
and a different treatment of high temperature partition functions, our 
strategy for calculating weak rates for A$\geq$65 is essentially the FFN 
approach. In this approach the rates are broken into two pieces: a
low part consisting of discrete transitions between individual levels,
and a high part involving the Fermi and GT resonances. This approach
is valid provided that discrete transitions between high lying levels
never dominant the rates. In addition, available experimental information for 
$\log ft$ values and level energies, parities, and spins is used. The inclusion 
of this much data for more than a hundred nuclei is a difficult task, and would
not be possible without the web-based nuclear structure databases 
(in particular those provided by nudat and the table of isotopes).

The simple FFN approach for estimating weak rates may be the most natural
framework for incorporating experimentally-determined nuclear
properties. Additionally, a virtue of using this semi-empirical
schematic approach is that we can easily see where the key nuclear
uncertainties are. Our work can then serve as a catalyst for more 
detailed follow up nuclear structure studies for important rates.

In the next section we present a discussion of the formalism for
calculating weak rates. In section III we discuss the assignment of
experimentally unknown matrix elements for allowed, discrete state
transitions. This discussion is based on the systematics of
experimentally observed weak transitions. In section IV we present a
simple approach for estimating the position of the GT resonances and
make some comparison with data and shell model calculations. The
calculation of rates in a high temperature environment is treated in section
V. We also discuss in section V some subtleties associated with the
proper partition function to be used in these calculations. Section VI
gives some comparisons of our estimates for the rates with those
provided by Langanke and Martinez-Pinedo.  Finally, we conclude with a
discussion of the results.

\section{Weak rates formalism}

The total decay rate for a nucleus in thermal equilibrium
at temperature $T$ 
is given by a sum over initial parent states $i$ and final parent
states $j$;
\begin{equation}
\label{partitionrate}
\lambda=\sum_i P_i \sum_j \lambda_{ij},
\end{equation}
where the population factor for a parent state $i$ with excitation 
energy $E_i$ and
angular momentum $J_i$ is
\begin{equation}
P_i={(2J_i+1)e^{-E_i/kT} \over Z}
\end{equation}
with 
\begin{equation}
Z=\sum_i(2J_i+1)e^{-E_i/kT}
\end{equation}
the nuclear partition function. Here $\lambda_{ij}$ is the specific weak transition
rate between initial parent state $i$ and daughter state $j$ and is formally given 
by 
\begin{equation}
\lambda_{ij}={\ln 2 \over (ft)_{ij}} f_{ij}
\end{equation}
where
\begin{equation}
{1 \over (ft)_{ij}} = {1 \over (ft)^{\rm F}_{ij}} + {1 \over (ft)^{\rm GT}_{ij}}
\end{equation}
and where the relation between ft-values and the appropriate Gamow-Teller
or Fermi matrix element is
\begin{eqnarray}
(ft)^{\rm GT}_{ij} & \approx & {10^{3.59}\over |M_{\rm GT}|^2_{ij}} \\
(ft)^{\rm F}_{ij} & \approx & {10^{3.79}\over |M_{\rm F}|^2_{ij}}.
\end{eqnarray}
Here the total Gamow-Teller matrix element between initial parent state
$|\psi^P_i\rangle$ and final daughter state $|\psi^D_f\rangle$ is
\begin{equation}
|M_{\rm GT}|^2_{ij}=|\langle\psi_f^D|\sum_n \sigma_n (\tau_{\pm})_n
|\psi_i^P\rangle|^2
\end{equation}
where the sum is over all nucleons. The GT strength satisfies the 
sum rule
\begin{equation}
S_{\beta^-}-S_{\beta^+}\equiv \sum_f |\langle \psi_f | \sum_n  \sigma_n (\tau_{-})_n | \psi_i^P\rangle|^2-|\langle \psi_f | \sum_n  \sigma_n (\tau_{+})_n | \psi_i^P\rangle|^2=3(N-Z).
\end{equation}
Similarily, the Fermi matrix element is 
\begin{equation}
\label{fermi}
|M_{\rm F}|^2=|\langle\psi_f^D|\sum_n(\tau_{\pm})_n|\psi_i^P\rangle|^2=T(T+1)-T_z(T_z- 1)=|N-Z|.
\end{equation}
This equation is derived by noting that $[T_+,T_-]=2T_z=(N-Z)$.

The phase space factors for $\beta^{\pm}$ decay are 
\begin{equation}
f_{ij}=\int_1^{q_n} w^2(q_n-w)^2G(\pm Z,w)(1-f_{\mp})(1-f_{\nu})dw
\end{equation}
and for electron/positron capture are 
\begin{equation}
f_{ij}=\int_{w_l}^{\infty} w^2(q_n+w)^2G(\pm Z,w)f_{\mp}(1-f_{\nu})dw
\end{equation}
where the upper(lower) signs are for electrons(positrons),
$q_n=(M_p-M_d+E_i-E_j)/m_ec^2$, and where $M_p$ is the nuclear mass of the
parent and $M_d$ is the nuclear mass of the daughter.
The threshold is $w_l=1$ for
$q_n>-1$ and $w_l=|q_n|$ for $q_n<-1$. The lepton occupation factors are
\begin{equation}
f_{\pm}=\left[ \exp{{U-U_{\rm F}^{\pm}}\over kT}+1\right]^{-1}
\end{equation}
where the (-) sign is for electrons, the (+) sign is for positrons, 
$U$ is the electron kinetic energy  and $U_{\rm F}^{\mp}$
the kinetic chemical potential (total chemical potential is here defined to include the
electron rest mass).
 The factor $G(\pm Z,w)$ is the coulomb wave correction factor
defined in terms of the Fermi factor F
\begin{equation}
G(\pm Z,w)\equiv {p\over w} F(\pm Z,w)
\end{equation}
as discussed in FFNI.

\section{Assignment of unmeasured, allowed transition 
matrix elements}

As in FFN, we break the rates $\lambda_{ij}$ into two distinct 
components: non-resonant discrete transitions between low lying 
levels, and transitions involving resonance states carrying total 
weak strength ($|M_{\rm GT}|^2+|M_{\rm F}|^2$) of order one or 
greater. 
Discrete state transitions between low lying states in the parent and 
daughter are important when the capturing lepton energies are too low
to reach resonance states in the daughter, or when the temperature is
too low to thermally populate parent levels with fast transitions to 
daughter states. To estimate matrix elements between low lying parent and
daughter levels, FFN adopted the simple prescription that all transitions
not forbidden by the selection rules have some average matrix element 
characteristic of a group of nuclei. The validity of this approach can
be addressed by looking at the systematics of matrix elements for 
experimentally observed $\beta^+$/ec and $\beta^-$ decays.

\begin{figure}
\epsscale{1.0}
\plotone{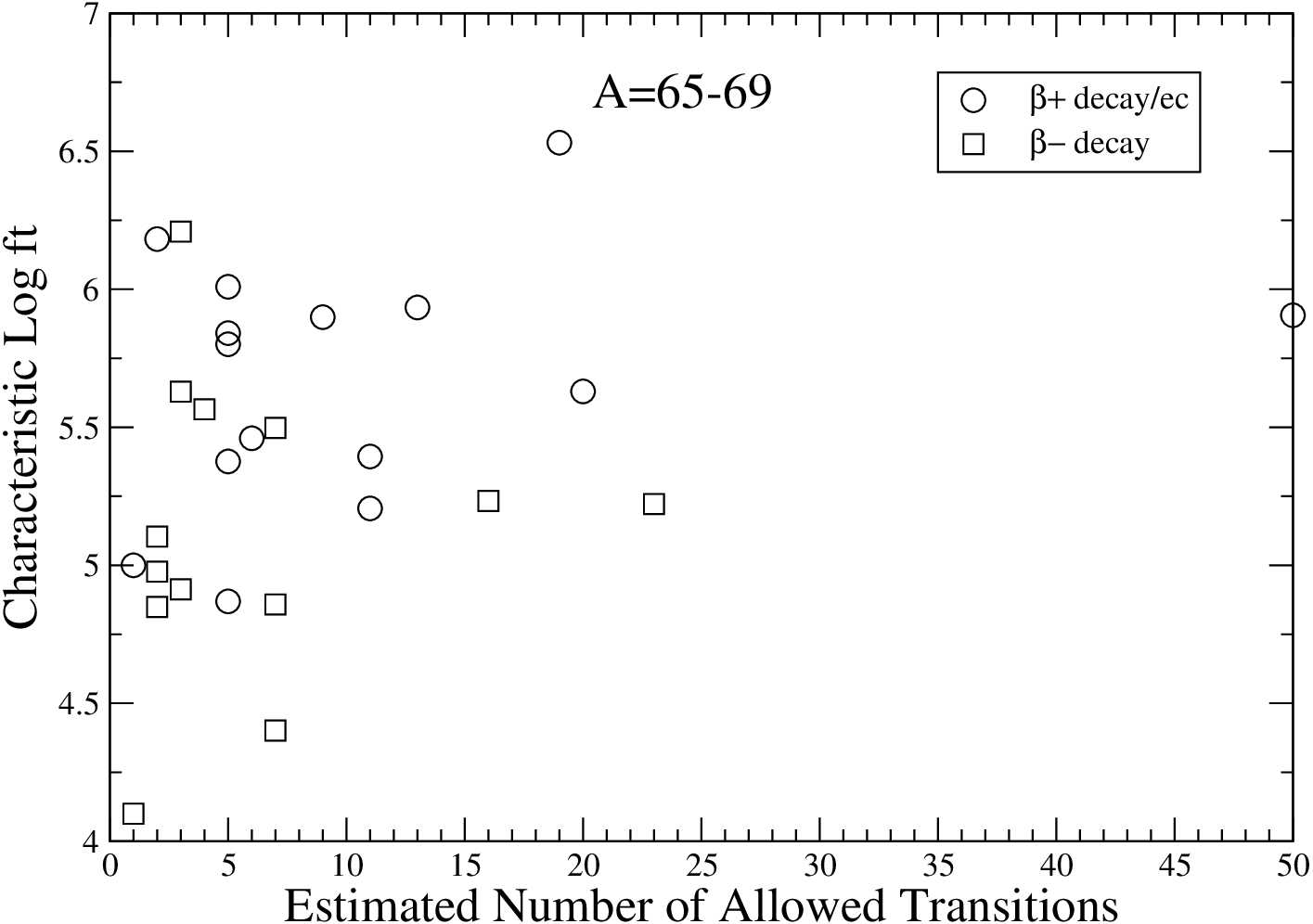}
\caption{ Systematics of 
experimentally-determined $\log ft$'s for nuclei with 65$\leq$A$<$70.
\label{char1}}
\end{figure}

\begin{figure}
\plotone{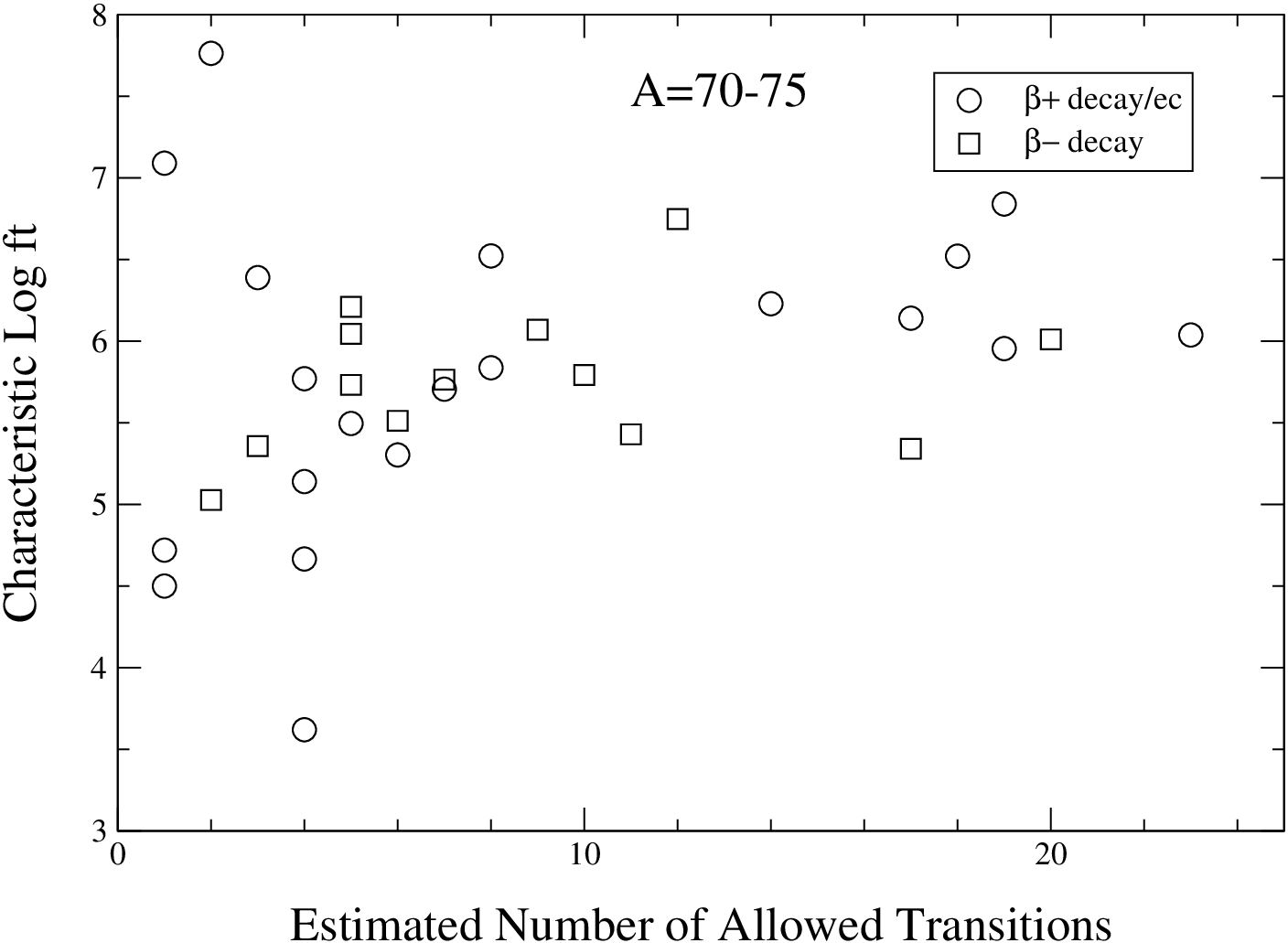} 
\caption{Systematics of 
experimentally-determined $\log ft$'s for nuclei with 70$\leq$A$\leq$75.
\label{char2}}
\end{figure}

\begin{figure}
\plotone{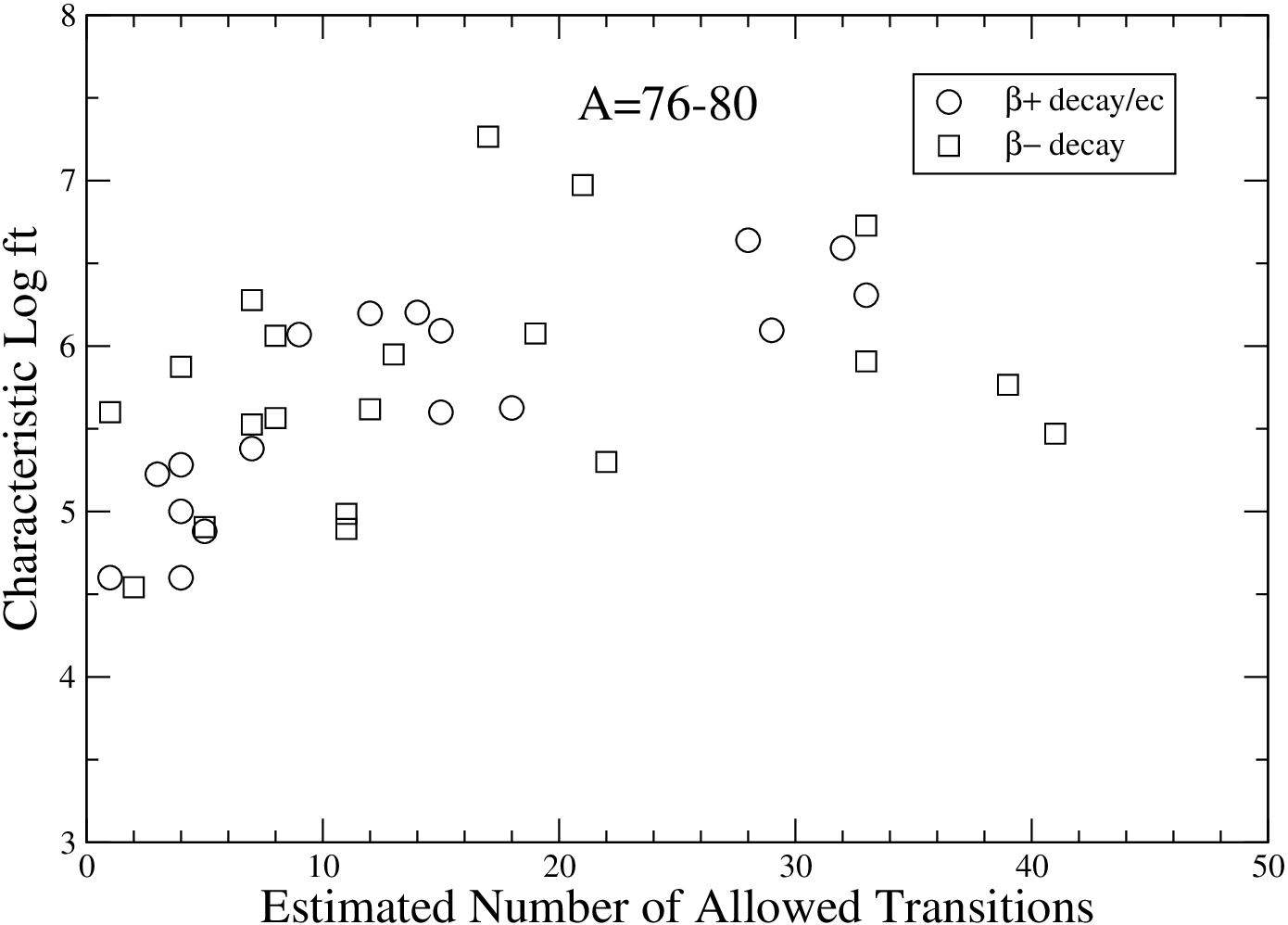} 
\caption{Systematics of 
experimentally-determined $\log ft$'s for nuclei with 75$<$A$\leq$80.
\label{char3}}
\end{figure}

Figures \ref{char1}, \ref{char2}, and \ref{char3} show the
characteristic $\log ft$ value for all nuclei in the mass range A=65-80
for which there is experimentally-determined weak decay
information. By "characteristic $\log ft$" we mean here the $\log ft$ value
obtained by assuming that all of the measured discrete strength is spread
uniformly over all states in the daughter for which the selection
rules do not forbid a transition and for which the Q-value is
positive. This estimated number of allowed transitions ($n_{\rm
allow}$) for each nucleus is shown on the x axis. The determination of
$n_{\rm allow}$ is difficult because of uncertainties in the angular
momentum and spin assignments of levels. Where the angular momentum of
a given level is uncertain we have adopted the middle value if more
than two possibilities are listed, and the largest value if only two
possibilities are given. Where the parity is listed, but uncertain, we
have adopted the tentative value. Transitions involving a level for
which no spin or parity information is given are labelled as
forbidden.

For illustration, consider the $\beta$ decay of $^{66}{\rm
Co}$. Experimentally it is observed that $^{66}{\rm Co}$ decays in
this channel to two states in $^{66}{\rm Ni}$, one with a $\log ft$ of
4.21 and the other with $\log ft=$4.75. By examining the experimentally
studied levels in $^{66}{\rm Ni}$ it is seen that there are 7 levels
that have spins and parities consistent with allowed decay from
$^{66}{\rm Co}$.  The characteristic $\log ft$ for $^{66}{\rm Co}$ is then
defined as $-\log_{10}((1/7)(10^{-4.21}+10^{-4.7}))$.  Our definition
of the characteristic $\log ft$ is rough in that it neglects the
experimental difficulties associated with measuring weak
transitions. In particular, uncertainties arising from the possibility
of the feeding of daughter states from higher lying states has not
been accounted for, nor has the difficulty of observing near-threshold
transitions. Nonethless, a case may be made from these figures that
the assumption of an average matrix element is a reasonable one,
particularly when several transitions are involved. For example, for
nuclei with $65 \le {\rm A} <70$, only one nucleus with $n_{\rm
allow}>2$ has a characteristic $\log ft$ differing from 5.4 by more than
1, and for nuclei in the range $70\le{\rm A} \le 80$ only 5 nuclei
with $n_{\rm allow}>2$ have a characteristic $\log ft$ differing from 5.7
by more than 1.

In this work we take a characteristic $\log ft$ of 5.4 for nuclei with
A$<$70, and a characteristic $\log ft$ of 5.7 for nuclei with
A$\geq$70. An alternative approach for estimating these matrix
elements would be to assign $\log ft$'s from a statistical
distribution. Our procedure should give a reasonable estimate of the
rates when several transitions contribute nearly equally to the rate.
However, our estimated rate is obviously subject to uncertainty when
only one or two
experimentally unknown transitions dominate. An important class of such
nuclei are the proton-rich even-even nuclei with nearly closed neutron
and proton sub-shells. These can have anomolously large $0^+\rightarrow
1^+$ GT transitions. In x-ray burst environments, with temperatures
$T\sim 0.25 {\, \rm MeV}$, the first $2^+$ excited state in these nuclei is
thermally populated. The thermal $\beta^+$ decay rate of the nucleus
depends sensitively on whether this $2^+$ state also decays with anomolously
large matrix elements. \cite{schatzphysrep} studied this question for a number
of proton-rich nuclei and found that the $2^+$ state decays more rapidly 
than the ground state when the ground state decays quickly. We incorporate
the results of the Schatz et al. (1998) shell model study for the four nuclei
($^{72,74}{\rm Kr},^{76,78}{\rm Sr}$) with fast $0^+\rightarrow 1^+$ 
transitions. For these nuclei we assume that the first $2^+$ state has a
strength distribution identical in shape, but with matrix elements twice as
large, as the strength distribution from the $0^+$ ground state. This gives
decay rates within $\sim 30\%$ of those calculated by Scahtz et al..

\section{The Fermi and Gamow-Teller Resonances}

The Fermi resonance $|F_i\rangle$ corresponding to a given state
$|\psi_i\rangle$ is generated by application of the isospin raising or lowering
operator, $|F_i\rangle = T_{\pm} |\psi_i\rangle$ (Eq. \ref{fermi}). 
The selection rules for Fermi transitions 
are $\Delta \pi =\Delta T =\Delta J=0$.  Because $\Delta T=0$
and because the ground state of a nucleus generally has the lowest
possible isospin, there typically is only non-zero Fermi strength for
transitions from a nucleus with greater isospin $T^>$ to a nucleus
with lesser or equal isospin $T^<$. Since the nuclear part of the Hamiltonian
($H_{\rm nuc}$) is isospin independent and the electromagnetic part is 
small in comparison, the resonance
generated by $T_{\pm}$ is narrowly concentrated about the IAS. The
excitation energy can be estimated from the difference in Coulomb
binding energy of the parent and daughter nucleus.  FFNI gives a useful
approximation for the excitation energy of the IAS in the daughter 
nucleus:
\begin{equation}
\label{ias}
E_{\rm IAS} =\Delta_p-\Delta_d \mp 0.7824\pm 1.728\min(Z_p,Z_d)/R
\end{equation}
where $R\approx 1.12A^{1/3}+0.78$ is the nuclear radius in fm, p and d
refer to the parent and daughter, respectively, $\Delta$ is the
atomic mass excess. The upper signs in the above equation correspond
to ${\rm (Z,N)\rightarrow (Z+1,N-1)}$ transitions for neutron rich
parents, while the lower signs correspond to ${\rm (Z,N)\rightarrow
(Z-1,N+1)}$ transitions for proton rich parents. Eq. \ref{ias} agrees
well with measured and shell model predictions for IAS energies.

The Gamow-Teller operator is 
\begin{equation}
GT_{\pm}=\sum_n {\bf \sigma} \tau^{\pm}(n),
\end{equation}
where the sum is over all nucleons $n$.  The collective GT resonance
state $|{\rm CGT_i}\rangle$ corresponding to a given parent state
$|\psi_i\rangle$ is given by application of the GT operator, $|{\rm
CGT_i}\rangle=GT|\psi_i\rangle$.  The selection rules for GT
transitions are $\Delta \pi=0$, $\Delta J=0,\pm 1$ no $0\rightarrow
0$, and $\Delta T=0,\pm 1$. The GT strength distribution is harder to
characterize because $H_{\rm nuc}$ is strongly spin dependent. Since
$[H_{\rm nuc},GT_{\pm}]\neq0$, the GT strength can be fragmented over
many daughter states. However, in practice the stellar weak rates are 
usually determined by the total strength and the centroid of the strength 
in excitation energy (provided that low lying discrete transitions 
are well accounted for).

The strength in the GT resonance was also estimated by FFN in a 
zeroth order shell model picture. In this picture the lowest shell orbitals 
are filled with nucleons, and the total strength is taken to be the sum of the
contributions from each pair of single particle orbitals:

\begin{equation}
\label{spstrength}
|M_{\rm GT}|^2=\sum_{\rm if} {{n_p ^i}{n_h ^f} \over 2j_f +1} 
|M^{sp}_{\rm GT}|^2_{\rm if}.
\end{equation}
Here i and f denote initial and final orbitals respectively, $n_p$ and $n_h$
denote the number of particles and holes in these orbitals, and 
$M^{sp}_{\rm GT}$ is the single particle matrix element connecting the initial
and final states. These single particle matrix elements can be found from 
angular momentum considerations and are shown in Table 1 of FFNII.

We follow FFN in using the single particle result (Eq. \ref{spstrength})
to estimate the total strength. Experimentally it is well established that 
the axial vector current is renormalized by a factor of $\sim (1/1.24)$ in 
nuclei. This results in a strength a factor of $(1/1.24)^2\approx 1/2$ smaller
than shell model calculations give. In addition, shell model calculations show that 
residual interaction-induced particle-hole correlations further reduce the total
strength by a factor of one to a few. Typically, these correlations are more important
for ${\rm GT^+}$ transitions, so that the additional quenching is larger for these transitions.
We adopt a quenching factor of 4 for ${\rm GT^+}$ transitions and 3 for ${\rm GT^-}$ transitions. These values 
for the quenching factors generally give strengths within a factor of two of 
more detailed strength determinations. When the quenched value for the strength is 
less than one, we assign a configuration mixing strength of one. As discussed
below, this is roughly consistent with the results of (n,p) experiments for 
${\rm GT^+}$ blocked nuclei.

FFN estimated the centroid of the GT resonance by considering a zeroth 
order shell model description for the spin flip part of the GT resonance.
This configuration was compared with the zeroth order shell model 
description of the daughter ground state and assigned an excitation energy
\begin{equation}
\label{egteqn}
E_{\rm GT}=\Delta E_{\rm s.p.} + \Delta E_{\rm pair} + \Delta E_{\rm ph}.
\end{equation} 

Here $\Delta E_{\rm s.p.}$ is the difference in single particle energies 
between the two states 
(daughter ground state and spin flip GT resonance state), 
$\Delta E_{\rm pair}$ accounts for the difference in pairing energy 
between the two states, and  $\Delta E_{\rm ph}$ (taken to be 2 MeV)
accounts for the effects of configuration mixing and particle hole 
repulsion. For $T^>\rightarrow T^<$ transitions the energy of the 
IAS in the daughter is added to eq. \ref{egteqn}.

For the assignment of the GT centroids we adopt a procedure close to
that outlined by FFN. However, as noted in the introduction, the FFN
approach misses some important systematics in the centroids of the
strengths as revealed by more recent experimental data and shell model 
calculations. One potential remedy for this is to do RPA or QRPA
calculations for the strength distributions for nuclei too heavy to be
studied via detailed shell model calculations. Such calculations have been done
for a number of nuclei by several different groups.  A simpler
approach is to approximately account for the effects of the
competition between the $k_{\sigma \tau}(\sigma_1 \cdot
\tau_1)(\sigma_2 \cdot \tau_2)$ and $k_{\tau} (\tau_1 \cdot \tau_2)$
terms in the nuclear Hamiltonian. The effect of this competition on
the centroid of the GT resonance is most easily seen in the Tamm
Dancoff approximation from the argument given by \cite{bertsch}. In
this approximation the $k_{\tau}$ and $k_{\sigma \tau}$ forces give
rise to a ${\rm GT^-}$ excitation energy scaling as 
\begin{equation} 
\label{bertschequation}
E_{\rm
GT^-} \approx E_{\rm IAS} + \Delta E_{s.o} + 2 (k_{\sigma \tau}
S_{GT^-}/3 -(N-Z) k_{\tau}). 
\end{equation}

Here $\Delta E_{s.o.}$ is the spin orbit splitting characteristic of
the single particle transitions for the 
${\rm GT^-}$ resonance. As an example of this equation, consider the case
of $^{54} {\rm Fe}$ and $^{56} {\rm Fe}$. The zeroth order $\beta^-$
strengths for these nuclei are $|M_{\rm GT}|^2=16.3$ ( $^{54} {\rm
Fe}$) and 22.3 ($^{56} {\rm Fe}$). For $k_{\sigma \tau} \sim (20/A)
{\rm MeV} \sim (2/3) k_{\tau}$, Eqn. \ref{bertschequation} implies
$(E_{GT^-} -E_{\rm IAS})_{^{56} {\rm Fe}}- (E_{GT^-} -E_{\rm
IAS})_{^{54} {\rm Fe}}
\approx -1 \,{\rm MeV}$.  Shell model calculations and (n,p) experiments for
these nuclei give this difference as approximately -2 MeV. As Fe
becomes more and more neutron rich $E_{\rm GT^-}$ approaches $E_{\rm
IAS}$ and eventually falls below it (although perhaps beyond the
neutron drip line for a nucleus as light as Fe).

A somewhat more sophisticated approach to incorporating the effects of 
the competition between the $k_{\tau}$ and $k_{\sigma \tau}$  forces is the
random phase approximation with a separable force. In this approximation the
GT$^+$ and GT$^-$ resonances become eigenstates of the Hamiltonian.
The energies of the resonances are approximately given by the roots of the
algebraic equation

\begin{equation}
\label{rpaminus}
{f_1 \over \epsilon_i-\epsilon+\Delta_{\rm s.o.}^-}+
{f_2 \over \epsilon_i-\epsilon}
+{f_3 \over -\epsilon_i+\epsilon+\Delta_{\rm s.o.}^+}
=-{3\over 2}{1 \over k_{\sigma \tau}} {1 \over S_{\rm tot}}
\end{equation}
(see \cite{gaarde} for an application of this equation 
to experimentally observed $\beta^-$ strengths, or 
\cite{rowe} for a more pedagogical discussion). In Eq. \ref{rpaminus},
$S_{\rm total}=S_{\rm GT^-}+S_{\rm GT^+}=3|N-Z|+2S_{\rm GT^+}$
is the sum of the strengths in the plus and minus directions, $f_1$
is the fraction of this strength in the ${\rm GT^-}$ spin-flip mode, 
$f_2$ is the 
fraction of this strength in the  ${\rm GT^-}$ non-spin flip mode, and $f_3$ is
the fraction of strength in the  ${\rm GT^+}$ direction. The spin orbit 
splittings $\Delta ^+ _{\rm s.o.}$ and  $\Delta ^- _{\rm s.o.}$ are 
the splittings appropriate for the spin flip transitions in the plus and minus
directions, $\epsilon_i=(\epsilon_{\pi}-\epsilon_{\nu})_i$ is the difference 
in the proton and neutron single particle energies for the levels involved 
in the transition. The quantity  $\epsilon_i$ is related to the 
energy of the IAS by $E_{\rm IAS}-\epsilon_i = 2 k_{\tau} |N-Z|$. The largest
root of Eq. \ref{rpaminus} corrresponds to the energy of the spin flip mode. 
Eq. \ref{rpaminus} reduces to Eq. \ref{bertschequation} in the limit
$f_1=1$.

Our approach for calculating the centroid of the 
 ${\rm GT^-}$ resonance is based on Eq. \ref{rpaminus}. The strengths
in this equation are estimated in the zeroth order shell model picture
described above. For consistency with FFN we take the spin orbit
splittings from Seeger and Howard (see the table in
\cite{thatprl}). When more than one spin flip transition contributes
to the strength we take the strength-weighted average. The parameter
$k_{\tau}$ can be estimated using measured IAS energies and estimates
for the particle-hole energies. Alternatively, an estimate for the IAS
energy (Eq. \ref{ias}) gives a relation between $k_{\tau}$ and
$\epsilon_i$. The parameter $k_{\sigma \tau}$ can be chosen to give
good agreement with shell model and experimental results for the 
${\rm GT^-}$
resonance in the fp shell. In this work we adopt $k_{\tau}=28.5/{\rm A}
{\,\,\rm MeV}$ and $k_{\sigma \tau}=23/{\rm A}{\,\,\rm MeV}$. These values
are close to those given in \cite{bertsch} and \cite{gaarde}.
It has previously been noted by a number of
authors (e.g. \cite{gaarde}) that a simple prescription can do a fair job of predicting the
centroids of the ${\rm GT^-}$ resonances. In table 2 we reaffirm this by
comparing the predictions based on Eq. \ref{rpaminus} with measured
and shell model results.

For the ${\rm GT^+}$ resonance, estimates based on a separable force are not well 
justified. Higher order particle hole correlations and correlations 
induced by other terms in the Hamiltonian play a more important role. 
Nonethless, the RPA result accounts for the spin orbit splittings 
and the systematics of the influence of the $k_{\tau}$ and
$k_{\sigma \tau}$ forces in an approximate way. We estimate the centroid
of the ${\rm GT^+}$ resonance from the equation analogous to Eq. \ref{rpaminus} (which
is found by reversing the role of the plus and minus transitions, and 
by setting $\epsilon_i=(\epsilon_{\nu}-\epsilon_{\pi})$ or alternatively 
by setting $\epsilon\rightarrow -\epsilon$ in Eq. \ref{rpaminus}).
We also add an additional term to account for the effects of correlations
missed by the separable force estimate:
\begin{equation}
\label{rpaplus}
E_{\rm GT^+}=E_{\rm GT^+,RPA} + \delta_{\rm ph}.
\end{equation} 
A value of 2 MeV for the empirical correction $\delta_{\rm ph}$ gives
good agreement with the results of experimental and shell model
studies of nuclei in the lower half of the fp shell. This is
demonstrated in Table 3.  

Our simple estimate for $E_{\rm GT^+}$
probably breaks down as the fp shell approaches being filled. Fortunately, 
for most nuclei in this case ($A>65$) 
the ${\rm GT^+}$ transition is nearly blocked (see
below). Likewise, for those proton rich nuclei with substantial amounts of
unblocked strength the electron capture Q-value is large, so again the 
rate is not terribly sensitive to assumptions about the centroid. For the few
proton-rich nuclei with modest ${\rm GT^+}$ strength, we can compare our
results to the more detailed QRPA calculations of
\cite{sarriguren1}. These authors performed QRPA studies for
even-even Ge, Se, Kr, and Sr isotopes. For the most part their
calculated electron capture half lives agree well with experimental
half lives. 

For $^{66}$Ge, $^{68}$Ge, and $^{70}$Se
\cite{sarriguren1} find that the strength distribution is not very
sensitive to which of two nearly degenerate shapes the nucleus
assumes, so that a direct comparison with our estimate is
sensible. For these nuclei our estimate for the centroid of the
resonance is lower than theirs by 1.5 Mev($^{66}$Ge), 1.8 MeV
($^{68}$Ge), and 2.5 MeV ($^{70}$Se).  For some of these nuclei it is
not clear which estimate is correct.  \cite{sarriguren1} show that
their method gives a centroid about 2 MeV higher than the experimental
centroid for $^{54}$Fe and $^{56}$Fe, and that there calculation for
$^{70}$Ge misses a modest amount of experimentally-determined low
lying strength. 

However, in at least one case ($^{70}$Se), our
assignment of the strength is at least 0.5-1 MeV too low, because it
results in too much strength within the electron capture window and
results in a half life about 5 times shorter than the experimentally
determined half life. (Note though, that where a half life is
available for a nucleus our calculations always agree with that half
life. We do not include resonance strength below the Q-value window for a
transition when experimental information is available.)
We will argue in the last section that the uncertainty in the
placement of the centroid of the ${\rm GT^+}$ strength is not very important
for nuclei in the mass range we are considering. 

As the ${\rm GT^+}$ strength decreases (i.e., as an isotope becomes more neutron
rich), Eq. \ref{rpaplus} eventually gives a negative excitation energy
for the resonance in the daughter. This typically happens when the
single particle estimate for the strength (Eq. \ref{spstrength}) is
less than about 5. In this case, the strength is dominated almost
entirely by configuration mixing. There are a few (n,p) studies of
such nearly blocked nuclei. \cite{vetterli} studied $^{70}{\rm
Ge(n,p)}^{70}{\rm As}$ and $^{72}{\rm Ge(n,p)}^{72}{\rm As}$.  The
experimentally-determined GT strength for $^{70}{\rm Ge}$ is
B(GT)=$0.84\pm 0.13$ or B(GT)=$0.72\pm 0.14$. The two different values
correspond to different ways of estimating the $\Delta {\rm L}=0$
component of the (n,p) cross section. The higher estimate comes from a
multipole decomposition (m-d) of the cross section, while the smaller
estimate is derived by approximating the cross section measured at
$5.8^{\circ}$ as the $\Delta {\rm L}=1$ component of the cross
section. For the m-d the strength distribution for $^{70}{\rm Ge}$ is
approximately flat up to a few tens of MeV in excitation energy in
$^{70}{\rm As}$. For the $5.8^{\circ}$ subtraction method, the
strength falls after about 6 MeV in excitation energy in $^{70}{\rm
As}$. For $^{72}{\rm Ge}$ B(GT)=$0.23\pm 0.05$ ($5.8^{\circ}$) or
B(GT)=$0.86\pm 0.14$ (m-d) and the strength distribution is roughly
flat or falls off after about 6 MeV depending on the background
substraction method. \cite{helmer} have studied $^{76}{\rm
Se}{\rm(n,p)}^{76}{\rm As}$. They find B(GT)=1.45 with a roughly flat
strength distribution for the m-d method, and B(GT)=0.35 with a
strength extending to about 6 MeV in $^{76}{\rm As}$ if the $6^{\circ}$
data is used to estimate the $\Delta {\rm L}=1$ component of the cross
section. In this work we adopt the procedure that when the RPA
estimate $E_{\rm GT^+,RPA}$ appearing in Eqn. \ref{rpaplus} is
negative with respect to the daughter ground state, the strength distribution 
is represented by a 1 MeV gaussian centered at 1.8 MeV. This is approximately
consistent with the experimental data analyzed using the $5.8^{\circ}$
($6^{\circ}$) background substraction method. Our prescription misses
the high lying strength estimated from the experimental data analyzed
using the multipole decomposition method. We will discuss in the last
section the uncertainties in the rates resulting from the unknown
strength distribution for these blocked nuclei.

\section{Thermal considerations}

The picture for Fermi and Gamow-Teller strengths outlined above
becomes more complicated at high temperatures on account of the
thermal population of parent excited states.  In evaluating the
contribution to the rates of transitions between low lying levels, we
use the same set of experimentally-determined levels discussed above.

As the temperature rises, the evaluation of Eq. \ref{partitionrate}
requires knowing the strength distributions of an impossibly large
number of states. For example, at temperature of $1\,{\rm MeV}$ the
partition function for an odd-odd nucleus with $A\approx 60$ is a few
hundred, and the mean excitation energy is $T^2 \rho_{\rm F}\approx
T^2 A/8\approx 7 \,{\rm MeV}$ (here $\rho_{\rm F}\approx a$ is the
level density at the fermi surface and $a\approx A/8$ is the level
density parameter).  The approximation traditionally (see FFNII) used
to make the problem tractable is the Brink approximation, which
postulates that the centroid of the Gamow-Teller strength distribution
corresponding to a parent state at excitation energy $E_i$ is shifted
up by an energy $E_i$ with respect to the centroid of the strength
distribution corresponding to the parent ground state. It is generally
assumed that the total strength remains the same for all transitions.
The validity of the Brink approximation has been investigated in some
detail by LMP.  They find that the approximation is good for the first
few low lying states for which they calculate strength distributions.

With the Brink approximation, the contribution of discrete state to
high lying resonance state transitions can be approximated. For
definiteness we assume in the following discussion that the parent
nucleus has isospin $T^>$, the daughter nucleus has isospin $T^<$, and
that the GT$^-$ operator acting on the parent generates states in the
daughter (and conversely). Each parent state ($|\psi_i^{\rm P}
\rangle$) has a corresponding Fermi resonance
($|F_i\rangle=T^-|\psi_i^{\rm P} \rangle$) and collective Gamow-Teller
resonance ($|CGT_i\rangle=GT^-|\psi_i^{\rm P}\rangle$) in the $T^<$ daughter.
With the Brink approximation, the Q-values for these transitions are 
independent of the excitation energy of the thermally populated 
parent state. The overall transition rate for these processes can 
be obtained brom a calculation of the ground state rate alone, but with 
the population factor of the ground state set to unity. This is part of the
``FFN trick''.

\begin{figure}
\plotone{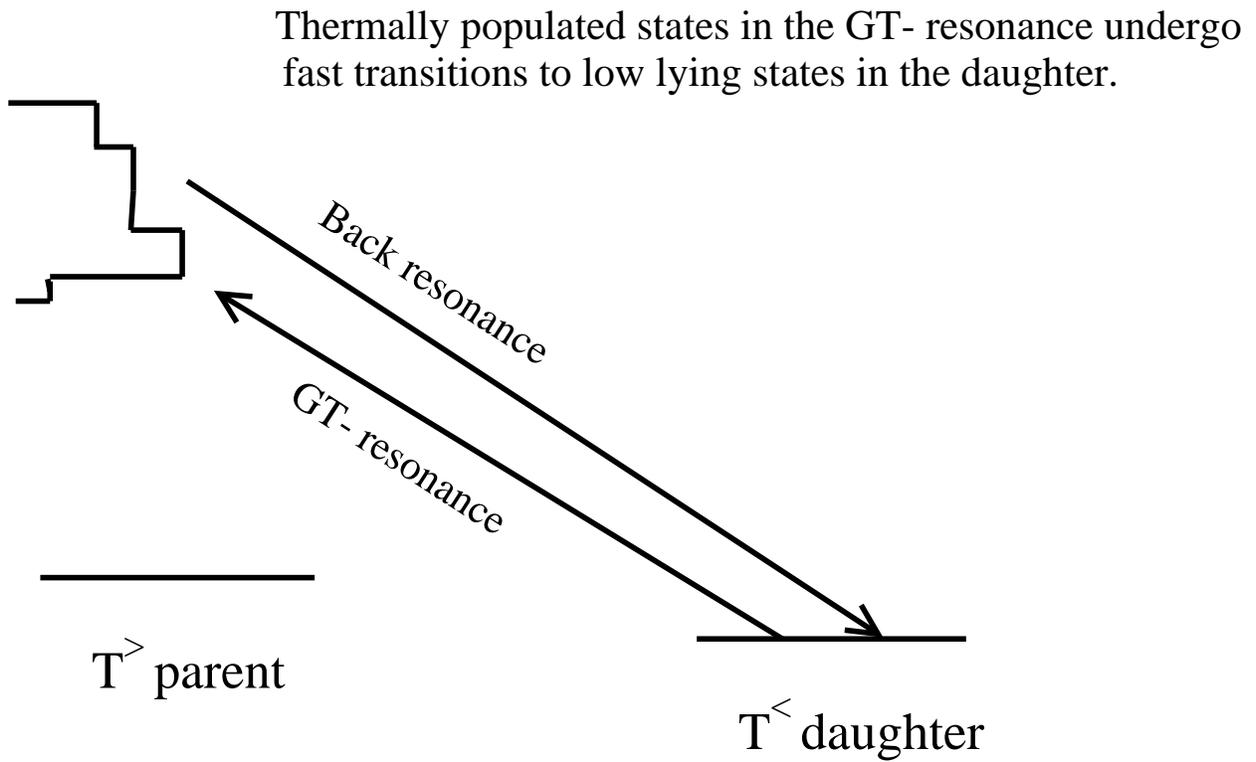}
\caption{Illustration of the position and energetics of 
discrete state-resonance transitions (back resonances). 
\label{backresfig}}
\end{figure}

Likewise, it is possible to thermally populate the collective Gamow-Teller
states in the $T^>$ parent ($|CGT_j\rangle=GT^+|\psi_j^{\rm D}\rangle$). These
thermally populated states can decay to the daughter with large overlap.
These ``back-resonances'' are shown in Fig. \ref{backresfig}.
Suppose for
a moment that the GT$^+$ resonance corresponding to a state in the
daughter is confined to a single state in the parent. In this case the
Brink approximation implies that for each state with excitation energy
$E_{\rm daughter,j}$ in the daughter, there is a corresponding GT$^+$
resonance state ($|CGT_j\rangle=GT^+|\psi_j^{\rm D}\rangle$)
in the parent with excitation energy $E_0+E_{\rm daughter,j}$.
Here $E_0$ is the excitation energy in the parent of the resonance state 
corresponding to the daughter ground state. Because in the Brink approximation
the Q-values and strengths for all the transitions $(E_0+E_j)_{\rm parent}
\rightarrow (E_j)_{\rm daughter}$ are the same, the rate contribution 
for each transition is also the same
and can be written as $\bar \lambda$. The contribution of the 
back resonances to the rates can be directly evaluated:

\begin{equation}
\label{backlambda}
\lambda_{\rm backres}={\bar \lambda \sum_i \exp(-E_{\rm parent,i}/T)(2J_{\rm parent,i}+1)\over Z_{\rm parent}} \\
\end{equation}
\begin{displaymath}
\approx {\bar \lambda \sum_j \exp((-E_{\rm daughter,j}+E_0)/T)(2J_{\rm daughter,j}+1)\over Z_{\rm parent}} \\
\end{displaymath}
\begin{displaymath}
=\bar \lambda  \exp(-E_0/T){Z_{\rm daughter} \over Z_{\rm parent}}.
\end{displaymath}
Eq. \ref{backlambda} was
first derived in FFNII and is sometimes referred to as the ``FFN trick''.

In this work we represent the resonance strength
distribution by a gaussian of nominal width $\sigma$. For the ${\rm GT^+}$ 
strength distribution we take $\sigma=1\,{\rm MeV}$
and for the ${\rm GT^-}$ strength distribution we take $\sigma=2\,{\rm MeV}$.
This approximation does not fairly represent the complex structure and variation
seen in real strength distributions. However, we will see in the last section 
that this simple approximation captures the features relevant for calculations 
of weak rates in a stellar environment.
The case where the resonance is spread over several states in
the parent can be treated similarily to the case where the resonance is concentrated
in a single state. If the resonance corresponding
to the ground state in the daughter is spread over states with
excitation energy $E_0,E_1,...E_n$ in the parent, with corresponding rates
to the daughter ground state given by $\bar \lambda_{0j}$, then
Eq. \ref{backlambda} becomes 
\begin{equation} \label{backrate2}
\lambda_{\rm back rate} \approx { \sum_j \bar \lambda_{0j} 
\exp(-E_{0j}/T) Z_{\rm
daughter} \over Z_{\rm parent}}.
\end{equation} 
This equation is
exact in the limit where the Brink approximation holds (i.e., the
strength distribution from every daughter state is identical to the
strength distribution from the daughter ground state except for an
overall shift in energy).  The interesting feature of
Eq. \ref{backrate2} is that if all of the Q-values for the n states
are comparable, then 
\begin{equation}
\label{eq24}
\lambda_{\rm back rate}\approx n \langle \bar 
\lambda_{0j}
\rangle \exp(-E_0/T) Z_{\rm daughter}/Z_{\rm parent}=\bar \lambda \exp(-E_0/T)
Z_{\rm daughter}/ Z_{\rm parent},
\end{equation}
 independent of n. Here the relation $n \langle \bar \lambda_{0j}
\rangle\approx \bar \lambda$ (where $\bar \lambda$ is the rate of
decay calculated assuming that the GT resonances are confined to
single states) holds because the total strength in the GT resonance is
independent of the number of states it is spread over.  In a sense
Eq. \ref{eq24}  is counterintuitive because one might expect that if the strength
is spread out over n states in the parent, the decay rate from the
parent should be a factor of 1/n smaller. However, in some cases the
strength distributions arising from different states in the daughter
must overlap (i.e. correspond to identical states in the parent) to
the extent that the level densities in the parent and in the daughter are
comparable. This is accounted for by the ratio of partition functions
appearing in Eq. \ref{backrate2}.

\begin{figure}
\plotone{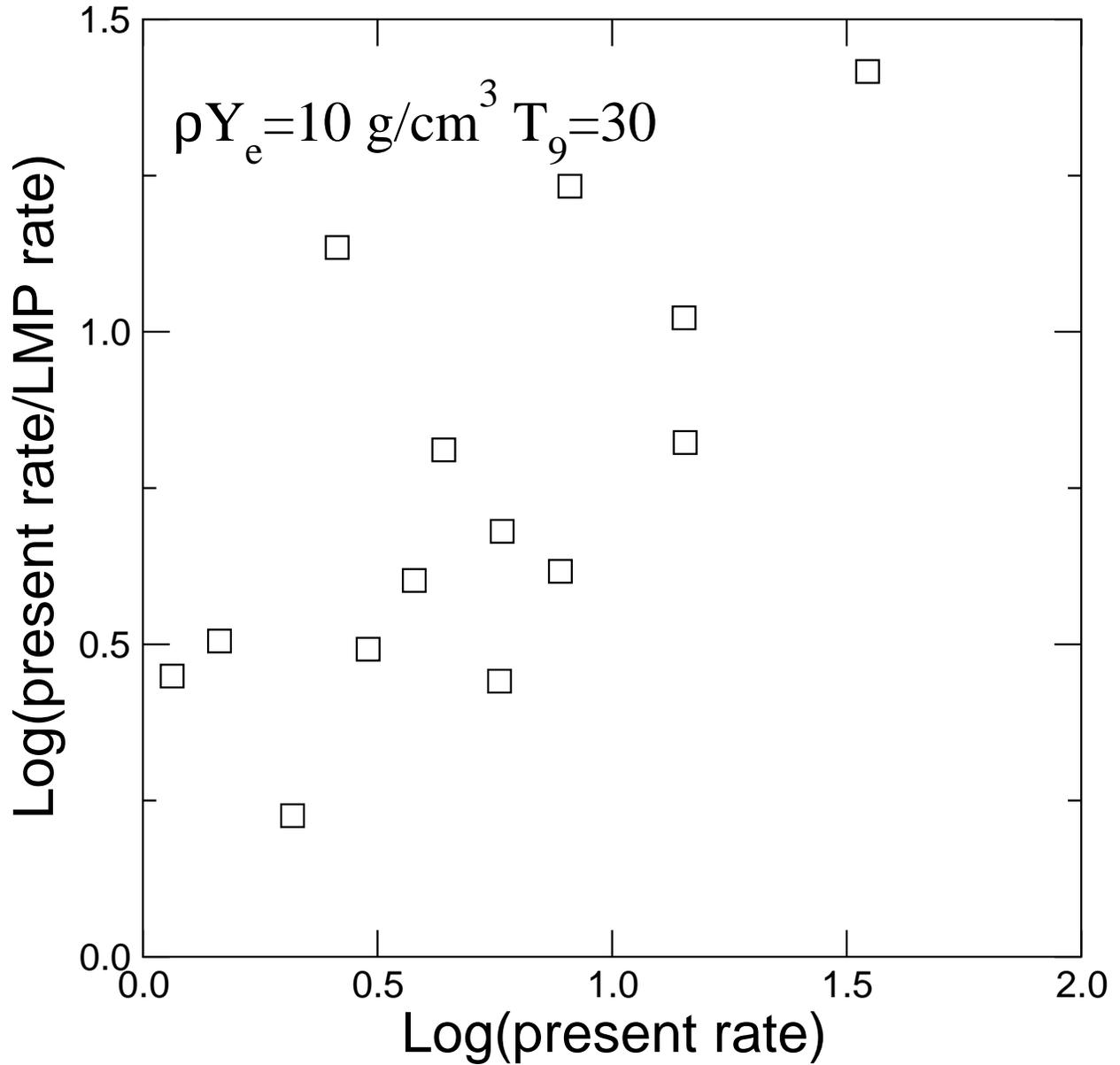}
\caption{Comparison of our $\beta^-$ decay rates with the LMP 
rates for nuclei in the mass range A=60-65. The comparison is made 
at $T_9=30$, $\rho Y_e=10{\rm g/cm^3}$. These conditions are artificial 
but serve to illustrate the influence of the treatment of the partition 
function on estimates of the weak rates. In all plots where a rate is shown
the rate is in units of ${\rm sec^{-1}}$.
\label{comparehigh}}
\end{figure}

In
principal, the back resonance contribution to the rates could be
calculated through a direct (e.g., shell model) calculation of the
strength distribution from the daughter states. In practice this
approach is not feasible because such a calculation would have to 
have fully converged final states corresponding to the strength 
distribution from many daughter states 
in order to distinguish which ones
are identical. Otherwise, an overcounting of the partition function
results, and the final rate estimate is a function of how many Lanczos
iterations are done. This line of reasoning indicates that at high
temperatures the current shell model-based calculations could systematically
underestimate the contribution to the $\beta$ decay rates from the
decay of the thermally populated ${\rm GT^+}$ resonance states by up to 
an order of magnitude. To show this, we plot in Fig. \ref{comparehigh}
a comparison between our calculated $\beta^-$ decay rates and those of
LMP at a temperature of $T_9=30$ and a density of $\rho Y_e =10{\rm g/cm^3}$.
These thermodynamic conditions are artificial, but serve the
purpose of illustration. When the $\beta$ decay rate is fast (and
insensitive to the finer details of the strength distribution), 
the discrepancy between the two calculations is about a factor of 
10. This is roughly the number of independent Lanczos states carrying
strength per daughter state in the calculations of LMP. From our simple
analysis it isn't clear that either set of rates is more reliable. However,
it is clear that differences in treatments of the partition functions 
can result in significant differences in estimates of the rates.

The trend of greater disagreement with faster decay rate seen in
Fig. \ref{comparehigh} arises from the competition of two factors.
For low decay rates, the Q-value of the decay (excitation energy of
the ${\rm GT^+}$ resonance with respect to the daughter ground state),
is typically small. In this case the width of the resonance enhances
the decay rate compared to the decay rate calculated from our
artificially narrow resonance. For faster decay rates, the Q-value is
large, the lifetime is relatively insensitive to the width of the
resonance, and the Lanczos-based calculation rate estimate is simply
suppressed by the overestimate of the parent partition function.

\begin{figure}
\plotone{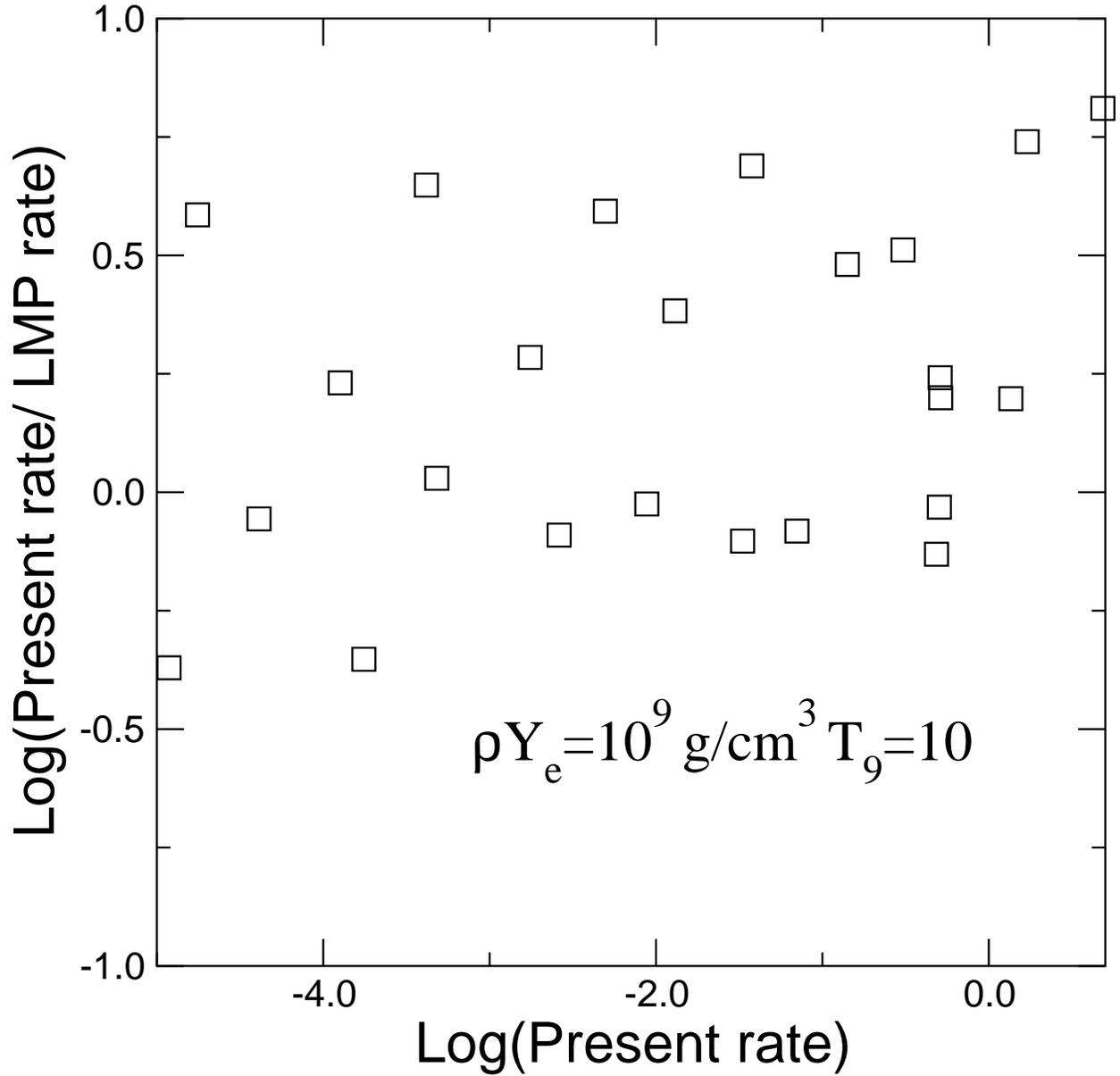}
\caption{Same as Fig. \ref{comparehigh} but at
$T_9=10$, $\rho Y_e=10^9{\rm g/cm^3}$, conditions approximating the
post Si burning degenerate core before collapse.\label{comparelow}}
\end{figure}

The effect, and the importance of consistency in partition functions in
general, can be illustrated by considering the conditions in the 
post-silicon burning, pre-collapse core of a massive star. There electron
capture proceeds on iron peak nuclei, driving them to a neutron excess where 
``reverse'' $\beta^-$ decay balances ``forward'' electron capture. For example,
\cite{aufder2} identify $^{64}{\rm Cr}$ as the endpoint nucleus where
the forward and reverse neutronization rates balance. In balanced conditions, 
partition functions are crucial rate and abundance determinants. We can compare our
rates with LMP as in Fig. \ref{comparehigh}, but now for $T_9=10$, $\rho Y_e=10^9
{\rm g/cm^3}$, roughly approximating immediately pre-collapse conditions.
This comparison is shown in Fig. \ref{comparelow}. Again there is a systematic
trend: the LMP rates are lower than ours by a factor of 4 on average with 
a fair scatter. This is smaller than the disagreement presented above for the
fastest $\beta$ decay rates, but still of potential significance given the dependence
of the initial ${\rm Fe}$ core mass on the electron fraction. Part of the
discrepancy between the rate estimates undoubtedly stems from differences in placement and
width of GT strength and other nuclear uncertainties. 

However, the partition effect outlined above likely plays a role as
well. In fact, these conitions ($T_9=10$, $\rho Y_e=10^9 {\rm
g/cm^3}$) are electron degenerate with Fermi energies $\sim 5{\,\,\rm
MeV}$, precluding significant $\beta$ decay for all but the nuclear
decay pairs with relatively large Q-values, just those where we argued
that the partition function-based uncertainties in rate estimates
could be large.

In evaluating the ratio $Z_{\rm daughter}/Z_{\rm
parent}$ in Eq. \ref{backrate2} we use the compilation of partition
functions from \cite{heger}. These partition functions include
experimentally-determined low lying levels and are supplemented at
higher excitation energies by a level density calculated from a back
shifted Fermi gas formula. At low temperatures the partition functions
of \cite{heger} agree well with the partition functions we calculate
for evaluating the rates between low lying levels. For temperatures
above 2 MeV we take $Z_{\rm daughter}/Z_{\rm parent}=1$. This is valid 
because the mean excitation energy at these temperatures is well above the 
pairing gap.

We do not claim that our partition function treatment is necessarily 
better than others, and it may well be inadequate in some conditions. In fact
there is a basic inconsistency in our rates: we include many states in our partition function
sums for which we include no weak interaction strength. Ideally we should 
include all states and all associated weak strength: only fully converged 
Lanczos and Monte Carlo calculations of weak strength and partition functions
currently do this. Failing to estimate partition functions and strength 
functions consistently can lead to inaccurate predictions of final ${\it equilibrium}$
parameters. In equilibrium, the hard won rates no longer matter and only the 
partition functions govern the final quantities of interest ($Y_e$, abundances,
etc.).

\section{Results and Discussion}

\subsection{A validity test: Comparison with Shell Model Based Rates 
for A=60-65}

Here we address the reliability of our calculated rates.  We have
shown in section 3 that with a simple prescription some gross
features of the strength distribution, in particular the total
strength and centroid of the distribution, may be estimated. However,
shell model and experimental results typically show a rich structure
in the strength distribution, with this structure varying markedly
from nucleus to nucleus (e.g. \cite{lan1}). Do the rates depend sensitively on the finer
features of the strength distribution? Or, alternatively, 
can a computationally simple method give a good estimate of the 
weak interaction rates? These questions can be addressed
by comparing our relatively simply derived rates with rates based on 
more detailed large dimension shell model calculations.

\begin{figure}
\plotone{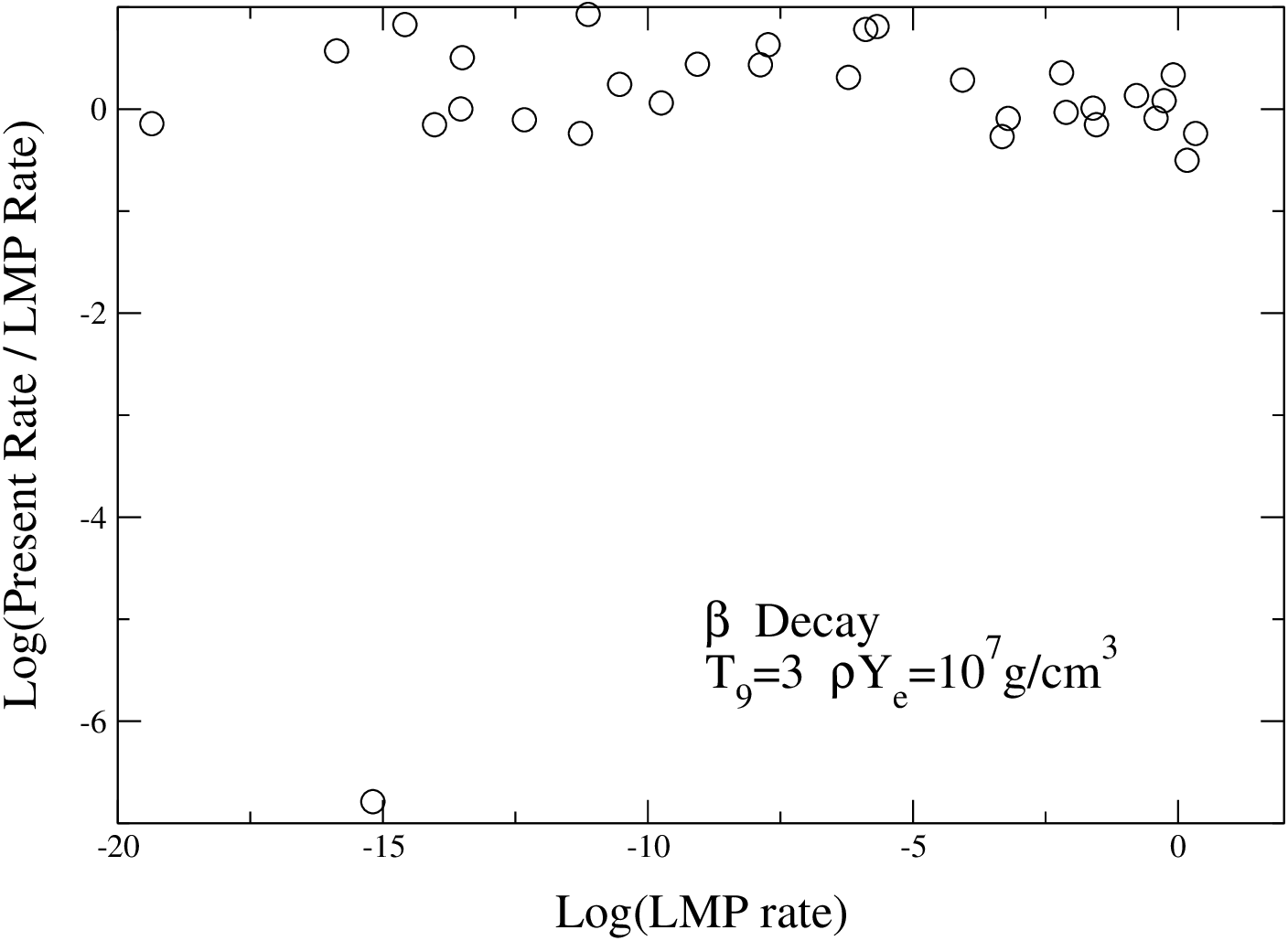}
\caption{Comparison of our beta decay rates calculated using a simple
estimate of the strength distribution with the more sophisticated
calculations of \cite{lan1} and \cite{lan2}. The different circles
correspond to different nuclei.  Comparisons for all nuclei in the
mass range A=$60-65$ for which \cite{lan2} provide rates are
presented.  Here $T_9=3$ and $\rho Y_e=10^7{\rm
g/cm^3}$.\label{comp1}} \end{figure}

\begin{figure}
\plotone{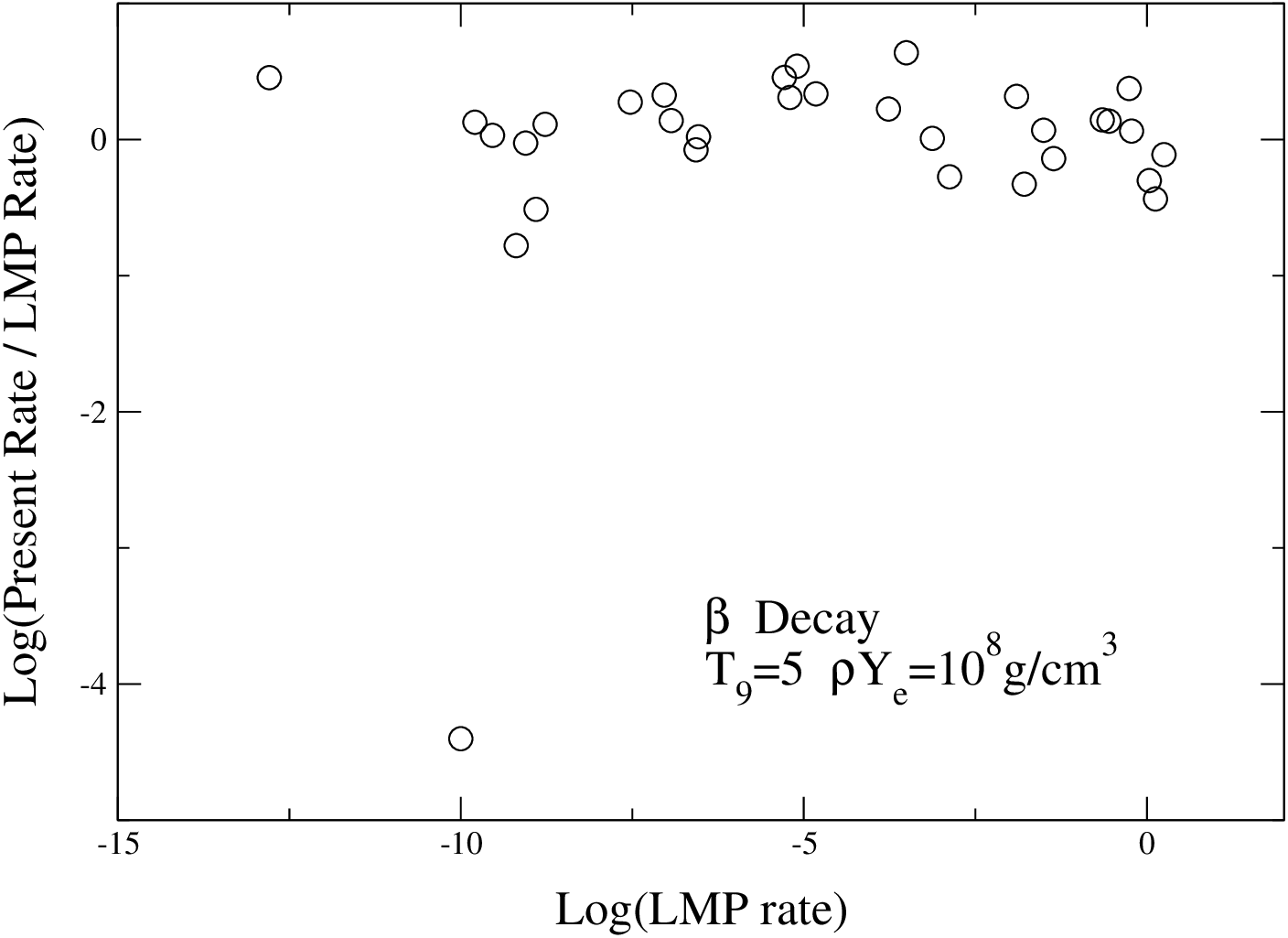} 
\caption{Same as
figure \ref{comp1} but with $T_9=5$ and $\rho Y_e=10^8{\rm g/cm^3}$\label{comp2}}
\end{figure}

\begin{figure}
\plotone{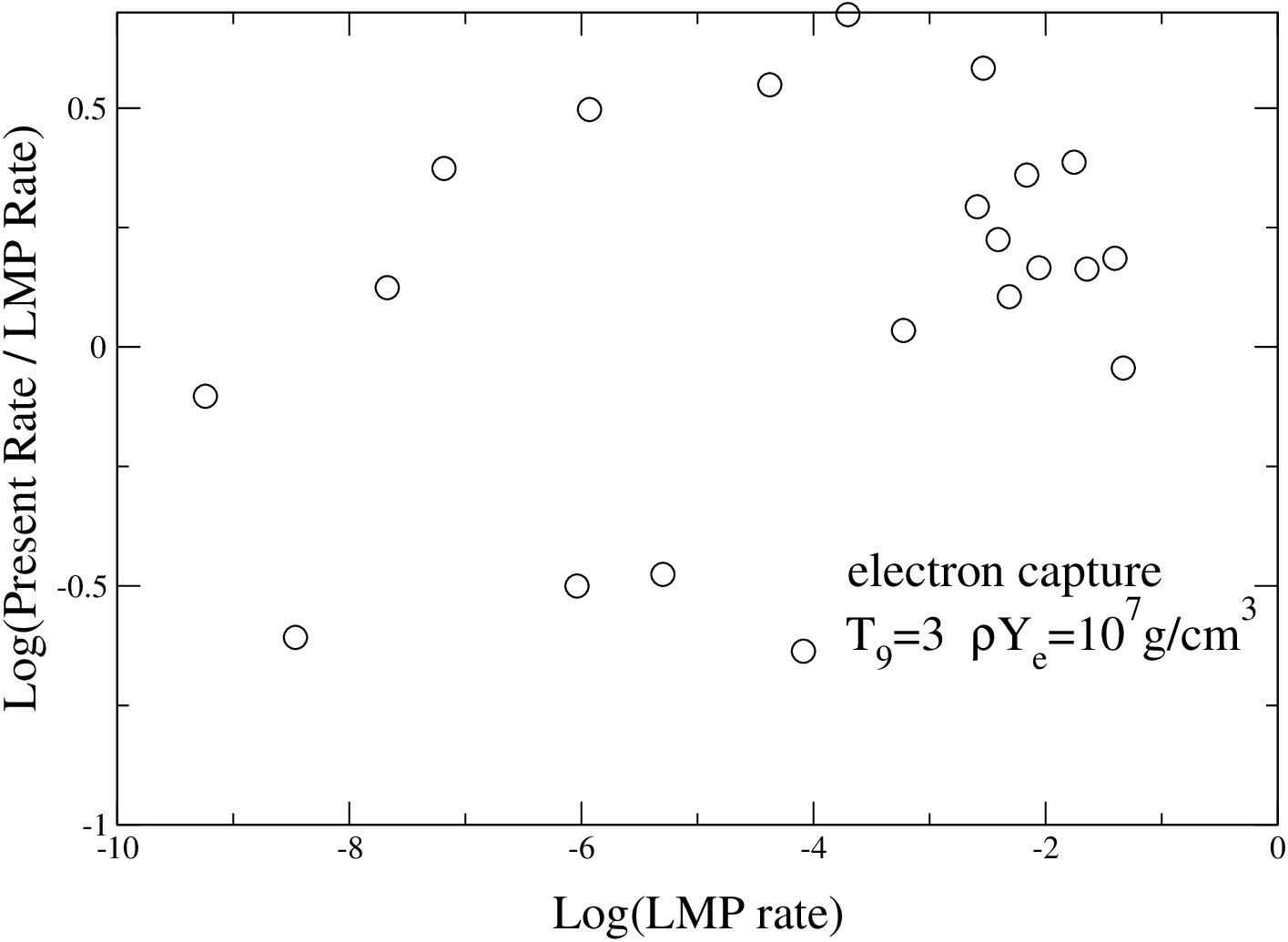}
\caption{Comparison of our electron capture rates with those of
Langanke and Martinez-Pinedo at
 $T_9=3$, $\rho Y_e=10^7{\rm g/cm^3}$. Only results for nuclei
that are estimated by LMP to have electron capture rates larger
than $10^{-10}{\rm sec^{-1}}$ are shown.
\label{comp4}}
\end{figure}

\begin{figure}
\plotone{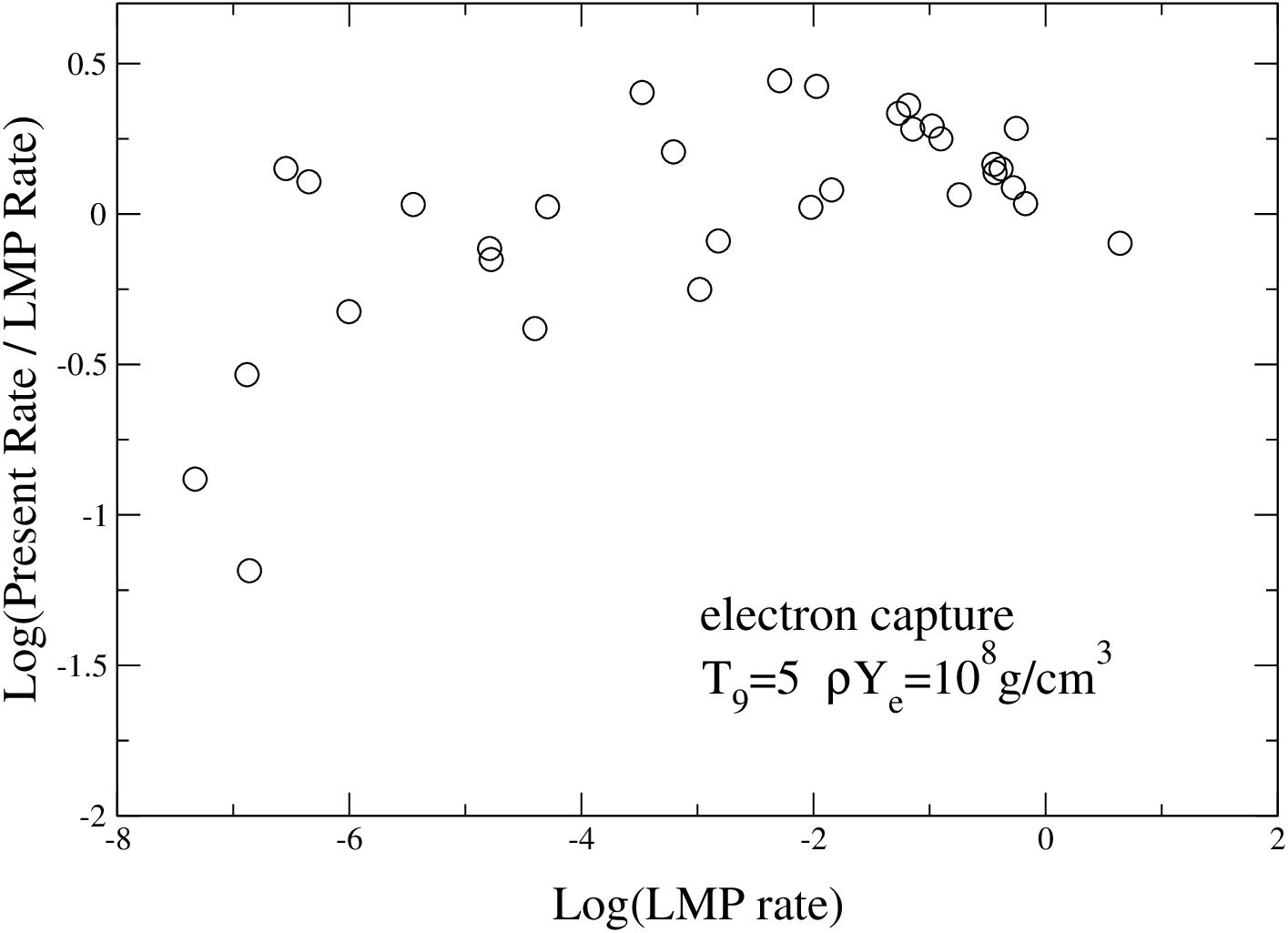}
\caption{Same as figure \ref{comp4} at $T_9=5$ and
$\rho Y_e=10^8{\rm g/cm^3}$.\label{comp5}}
\end{figure}

\begin{figure}
\plotone{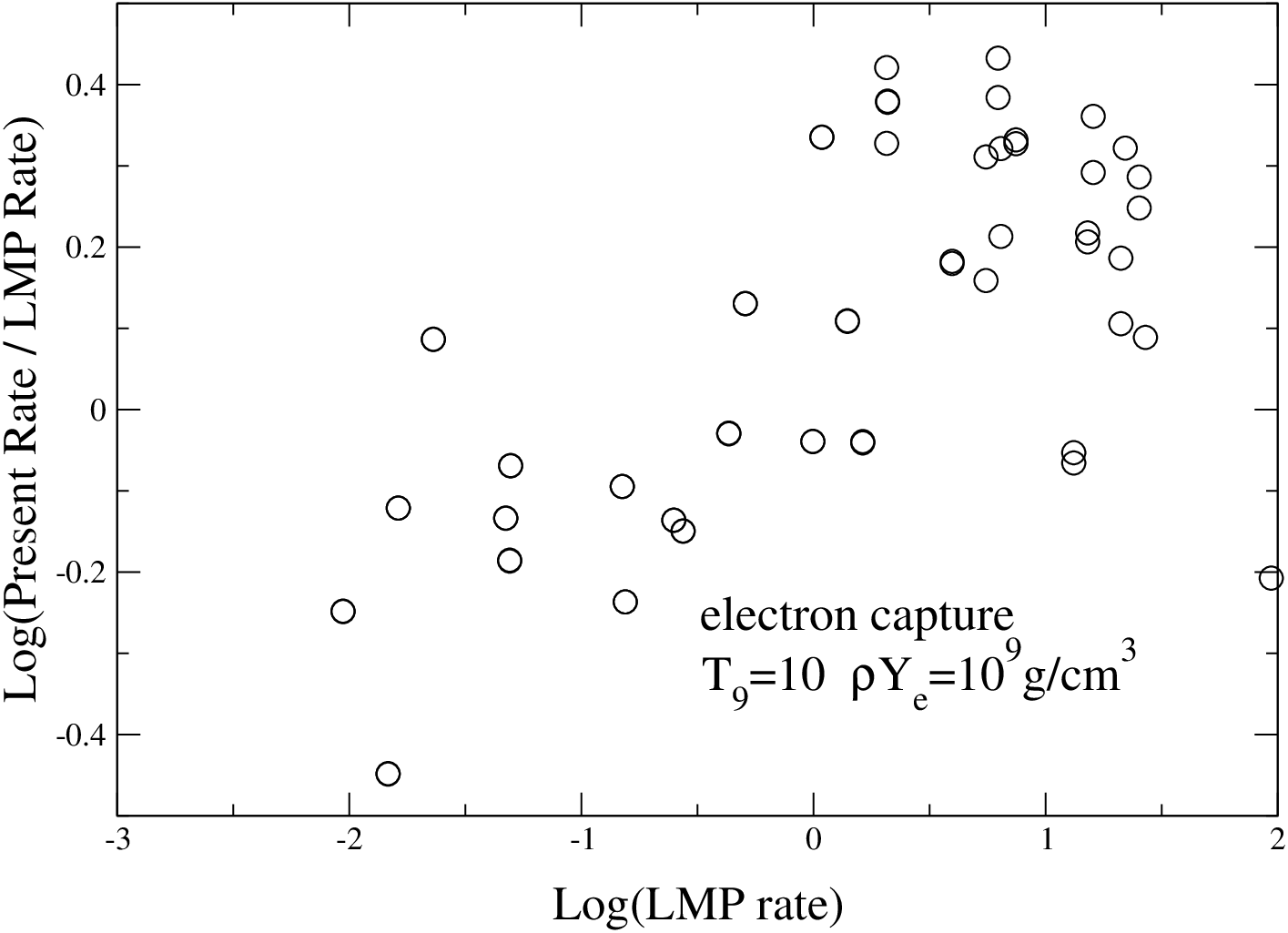}
\caption{Same as figure \ref{comp4} at $T_9=10$ and $\rho Y_e=10^9{\rm g/cm^3}$.
\label{comp6}}
\end{figure}

In figures \ref{comp1} and \ref{comp2} we show the log of
the ratio of our calculated beta decay rates to those calculated in
\cite{lan2} at several temperature/density points relevant for stellar
collapse, and for all nuclei in the range A=60-65 for which LMP
calculated rates. The horizontal axis in this figure is the log of the
beta decay rate calculated by LMP. We have presented the comparison in
this way because nuclei with very small rates will not be so important
in determining the evolution. The comparison is generally remarkably
favorable, with typical results differing by less than a factor of
two. Figures \ref{comp4},\ref{comp5} and \ref{comp6} are the same as
the previous three figures, but now electron capture rates are being
compared. Again, the comparison is favorable, the rates as calculated in the
two schemes being within a factor of three or so. This is surprising 
given the potential sensitivity of electron capture rates to the placement
and the width of the Gamow-Teller distribution.

The fact that a simple prescription and rough estimates of the
strength do a reasonably good job in getting the rates relevant for
stellar collapse is not to say that all rates are accurately
determined at all temperatures and densities with a simple model. Some
of our calculated rates deviate significantly from more detailed
calculations. This typically occurs for $\beta^-(\beta^+)$ decay when
thermal population of the states in a collective Gamow-Teller
resonance results in an exponential sensitivity, or for lepton capture
when the maximum lepton energies are just on the edge of being able to
reach the resonance in the daughter. Typically this exponential
sensitivity is accompanied by a very small rate, so that it is not
very important for stellar evolution.
\subsection{Examples of some rates important in X-ray burst environments}

X-ray bursts arise from the thermonuclear burning of hydrogen and
helium accreted onto the surface of a neutron star in a binary system
(see \cite{woosrp}). Characteristic temperatures during burning are a
few hundred keV, and characteristic densities are $\rho Y_e\approx
10^6 {\rm g/cm^3}$. The creation of nuclei heavier than ${\rm
A}\approx 40$ occurs via the {\it rp-}process, in which nuclei undergo
$({\rm p,\gamma})$ reactions until they approach the proton drip line
and/or $\beta^+/{\rm ec}$ decay intervenes. The time for burning along
the {\it rp-}process path is set to some extent by the $\beta^+/$ec
lifetime of a few important waiting point nuclei. Here we present a
few examples of important rates.

Three important waiting point nuclei with A$<$80 are the even-even
(proton bound) nuclei $^{64}{\rm Ge}$, $^{68}{\rm Se}$, and $^{72}{\rm Kr}$ \citep{schatz}.
For each of these nuclei the single proton capture daughter (Z+1,N) is unbound,
while the next heaviest isotone (Z+2,N) is bound. At low temperatures
$(T_9 \lesssim 1.5)$ the weak rate most important for determining flow towards
the valley of $\beta$ stability is $\lambda_{\beta^+}$(Z,N). For higher temperatures
an equilibrium between (Z,N) and (Z+2,N) is reached, so that $\lambda_{\beta^+}
(Z+2,N)$ is also important. As a typical example we discuss $^{72}_{36}{\rm Kr}$ and 
it's two-proton capture daughter $^{74}_{38}{\rm Sr}$. 

The first $2^+$ excited state of $^{72}{\rm Kr}$ lies at $\approx 700\, {\rm keV}$
and is not significantly populated for temperatures $T_9 < 2.5$. The ground state lifetime 
of $^{72}{\rm Kr}$ is experimentally-determined. The low lying strength distribution in 
the 
daughter ($^{72}{\rm Br}$) has also been measured. With the ground state lifetime and
low lying strength distribution measured, the only missing piece of information 
is the strength distribution at excitation energies too high to 
be experimentally observed. Our rough estimate (Eq. \ref{rpaplus}) places a sizable
portion of the resonance strength within the Q-value window for the decay. In order not
to conflict with the experimentally-determined lifetime we push this strength 
up to Q=0 in our calculation. 

\begin{figure}
\plotone{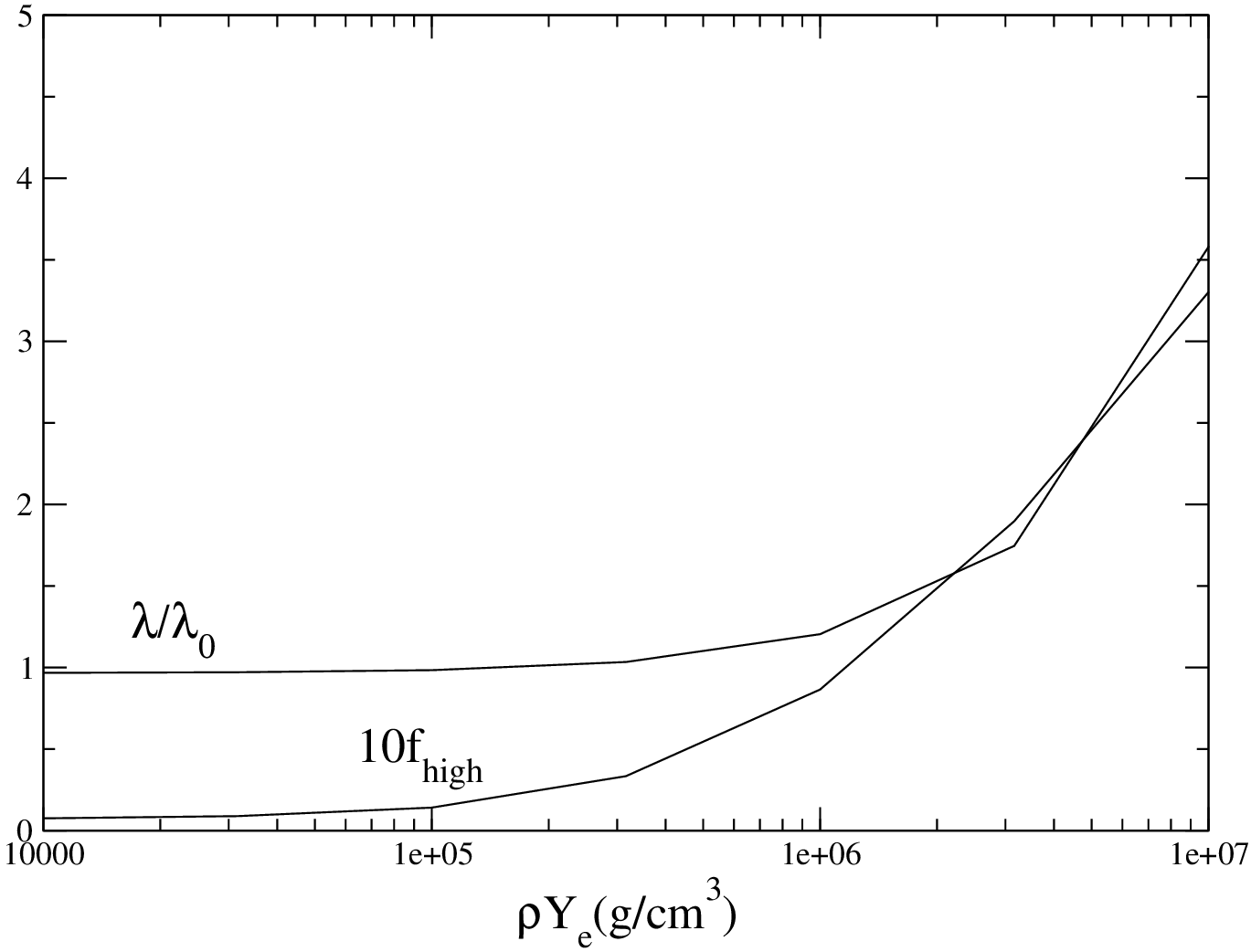}
\caption{Influence of high lying resonance strength on the total rate
($\beta^++ec$) for $^{72}{\rm Kr}$. In this figure $f_{\rm high}$ is
an estimate of the fraction of the total rate arising from transitions
involving experimentally unmeasured, high lying resonance strength.
The upper curve, labelled $\lambda/\lambda_0$, shows the ratio of the
total rate to the $\beta^+$ decay rate at zero temperature.
\label{kr72.eps}}
\end{figure}

To quantify the uncertainty in the rates arising from the high lying resonance
strength we plot in Fig. \ref{kr72.eps} the total rate (${\beta^+ + {\rm ec}}$)
for $^{72}{\rm Kr}$. Also included in this figure is the fraction of the total 
rate coming from Q=0 and above. The figure shows that the experimentally-determined
lifetime is sufficient for a determination of the thermal decay rate at the
$10\%$ level for $\rho Y_e < 10^7 {\rm g/cm^3}$. At $\rho Y_e=10^7 {\rm g/cm^3}$, 
our simple estimate shows the high lying strength accounting for 
$\sim 20\%$ of the total weak rate. Because our method puts essentially all
of the strength at Q=0, and very little above Q=0, it is unlikely that 
we have underestimated the contribution to the rate from the high lying resonance
strength. 
 
The decay of $^{74}{\rm Sr}$ is more easily calculated because the
ground state (or a low lying excited state) of the odd-odd N=Z
daughter ($^{74}{\rm Rb}$) is the IAS of the ground state of
$^{74}{\rm Sr}$.  The large matrix element and Q-value ($\sim 10
\,{\rm MeV}$) for the Fermi decay mean that electron capture cannot
compete with $\beta^+$ decay in x-ray burst conditions for this
case. It is difficult to reliably estimate the contribution to the
$\beta^+$ rate from ${\rm GT^+}$ transitions, but a reasonable
estimate is that the diffuse ${\rm GT^+}$ strength only decreases the
ground state lifetime by at most $10-20\%$. 

By mirror symmetry, a low
lying thermally populated state of $^{74}{\rm Rb}$ corresponds to the
IAS of the next even-even $\beta^+$ nucleus ($^{74}{\rm Kr}$), so that
the chain $^{74}{\rm Sr}\rightarrow ^{74}{\rm Rb} \rightarrow
^{74}{\rm Kr}$ is fast and dominated by Fermi transitions. The decay of the 
odd-odd N=Z nucleus ($^{74}{\rm Rb}$ in this example) should is 
calculated as the decay of a thermally populated back resonance as
discussed above. This is because only those parent states with an IAS in the
even-even daughter nucleus decay via the Fermi transition. Since the partition
function of the odd-odd parent increases more rapidly with temperature
than the partition function of the even-even daughter, the decay rate decreases
rapidly with increasing temperature.
Another set
of nuclei important for the {\it rp}-process with decays dominated by the
Fermi transitions are those nuclei with ${\rm (Z=N+1,N=even)}$. These
have an IAS near or at the ground state of the $\beta ^+$ daughter.

Generally, for a given ground state lifetime, the electron capture rate 
is small and insensitive to the placement of strength within the Q-value window
as long as the initial state electron energy is small compared to the 
energy of the decay, i.e.
\begin{equation}
\langle E_e \rangle < ({\rm M_p-M_d+E_i-E_f}).
\end{equation}
In the x-ray burst environment $\langle E_e \rangle < 0.5\,{\rm MeV}$,
while the Q-values for the proton-rich nuclei are typically greater
than $4\,{\rm MeV}$, so the thermal lifetime is reliably estimated
from the ground state lifetime.  By contrast, nuclei closer to the
valley of stability have smaller Q-values and therefore the lifetimes
are nearly entirely determined by thermal electron capture.

\begin{figure}
\plotone{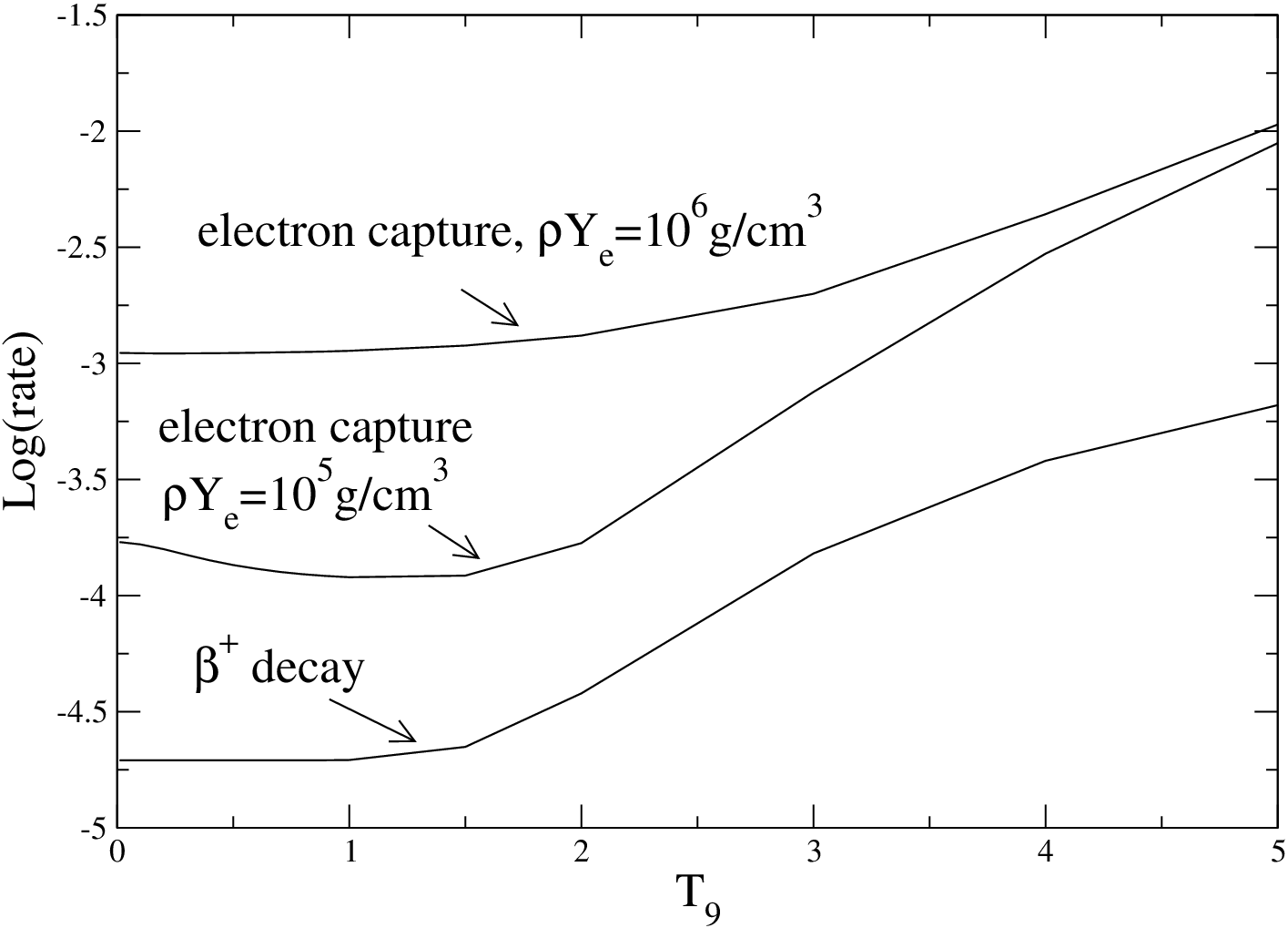}
\caption{$\beta^+/{\rm ec}$ rates for $^{66}{\rm Ge}$.The increase of the
$\beta^+$ rate with temperature arises from the fast decay of the
thermally populated first $J^{\pi} =2^+$ excited state. Unlike the $0^+$ ground
state, the $2^+$ state can have allowed transitions to several $2^+$
and $3^+$ daughter states. The decrease in the electron capture rate
at low temperatures is likely an artifact of our calculation and
arises because the $0^+$ ground state has substantial ${\rm GT}$
strength at high daughter excitation energies. As the daughter level
structure is poorly known at these high excitation energies, our
calculation does not include estimates of the allowed matrix elements 
for the decay of the first excited $2^+$ state to those states.
\label{66ge.fig}}
\end{figure}

For example, consider $^{66}{\rm Ge}$, a nucleus important in steady
state burning on and/or near neutron star surfaces. This nucleus has a
Q-value of only $\sim$2 MeV and a fair portion of the strength lies at
the upper end of the Q-value window. Consequently thermal electron
capture dominates over positron decay for $\rho Y_e \gtrsim 10^5 {\rm
g/cm^3}$. This is shown in figure \ref{66ge.fig}.

\subsection{Weak rates in the late time pre-supernova star}

Here we discuss some of the systematics of the weak rates in the
hot and electron-degenerate core during the $\sim10^4$s before
core bounce. Most of the nuclei with A$>$65 present in the core will
be blocked or nearly blocked to $GT^+$ transitions in the zeroth 
order shell model picture. Our calculation of the electron capture
rates for these nuclei is based on an estimate of the $GT^+$ configuration
mixing strength. Forbidden transitions and transitions allowed by the
thermal unblocking of strength will also contribute to the electron capture
rates. It is useful to estimate how large the configuration mixing strength
must be in order to justify the neglect of these latter types of transitions.

Thermal unblocking refers to the population of parent excited states
that have zeroth order shell model configurations consistent with
allowed transitions to states in the daughter. Assume that these
parent states comprise a fraction $\delta Z$ of the total partition
function. Then, an effective thermal unblocking matrix element is
$|M_{\rm TU}|^2\approx (\delta Z/Z) |M_{\rm s.p.}|^2$.  Here $|M_{\rm
s.p.}|^2\approx 1-3$ is a typical single particle allowed transition
matrix element. An accurate estimate of $\delta Z$ is very difficult
and an important open issue. Fuller 1982 parametrizes $\delta Z\approx Z
exp(-E^*/T)$, with $E^*$ the excitation energy of the lowest parent
excited state with an allowed $GT^+$ transition in the zeroth order
shell model picture. With this schematic notation, thermal unblocking
can compete with a configuration mixing strength of $0.1-1$ if
$E^*/T\lesssim 2$. For a typical $E^*\sim 5\,{\rm MeV}$, then, thermal
unblocking can be neglected for $T\lesssim 2{\rm \, MeV}$.  for
$T>2{\rm\, MeV}$ the thermal population of the $GT^-$ resonance states
(which we do include in our calculations) also becomes important.

Forbidden transitions become important as the electron chemical
potential increases and the wavelength of the leptons involved in an
electron capture event become small enough to probe structure in the
nucleus. Again following the convention of Fuller 1982, the
contribution to the electron capture rate from forbidden transitions
can be written as $\lambda_{\rm for}\sim |M_{\rm for}|^2 f_{\rm
for}(E_{\rm for},Q, \mu_e)$. Here $|M_{\rm for}|^2\sim 10-20$ is
roughly the number of protons in the fp shell multiplied by a typical
single particle first forbidden matrix element.  The unique
first forbidden phase space factor $f_{\rm for}$ depends on the
centroid in energy of the forbidden strength distribution $E_{\rm
for}$, the parent-daughter mass difference $Q$, and the electron
chemical potential $\mu_e$. For a typical $Q=10{\rm\, MeV}$, forbidden
transitions compete with a low lying configuration mixing strength of
$\sim 1/2$ for $\mu_e=31{\rm \, MeV}$ if $E_{\rm for}=5{\rm \, MeV}$. If the
centroid of the forbidden strength lies instead at $\sim 10{\rm \, MeV}$ above
the daughter ground state, then forbidden transitions do not become important
until $\mu_e>37{\rm \, MeV}$.

\begin{figure}
\plotone{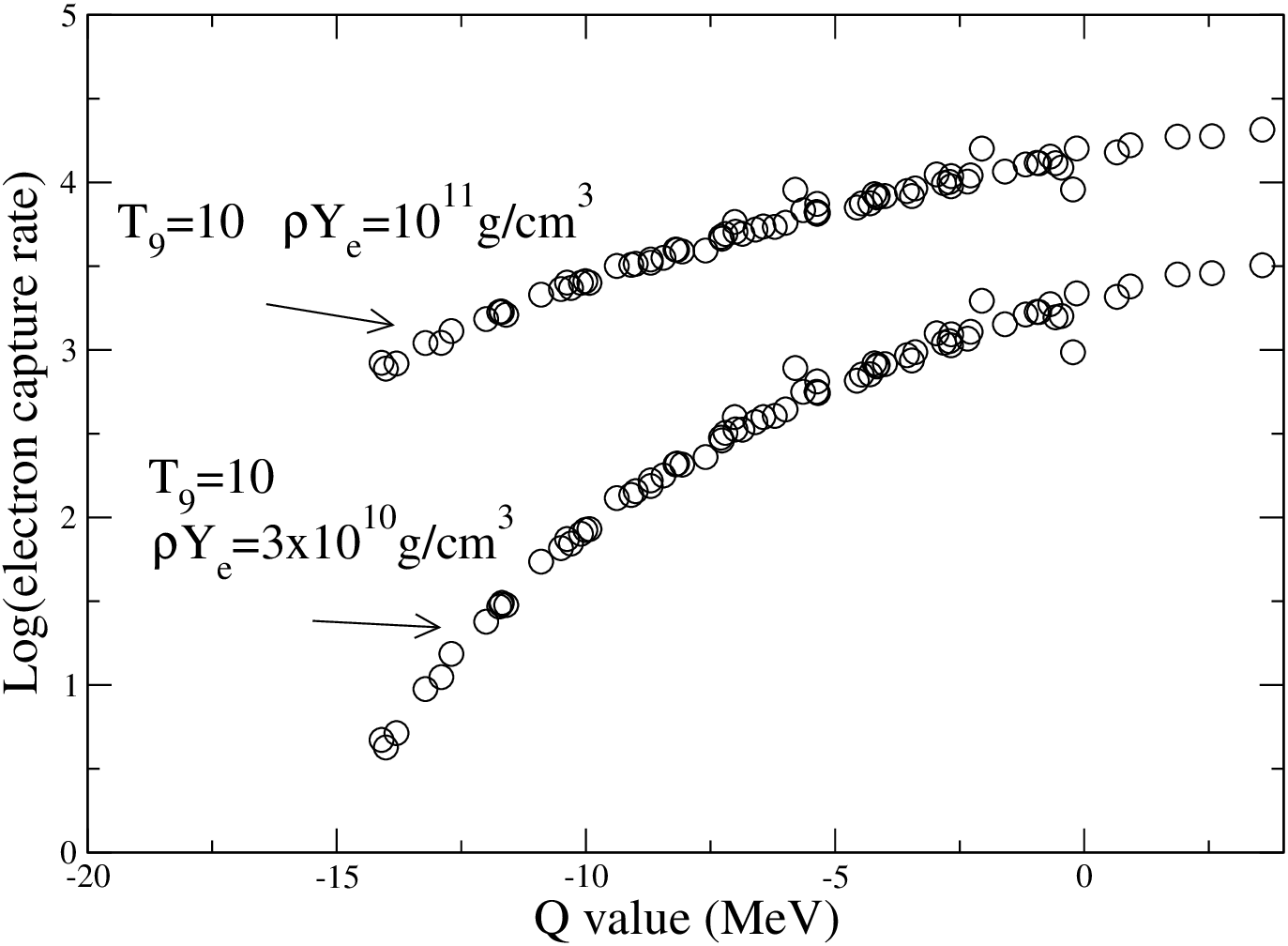}
\caption{Electron capture rates for nuclei in the mass range
A$<65\leq80$ and for which (N-Z)/A$>$0.1 as a function of the Q-value
 (defined here as the parent daughter atomic mass difference).
These rates have a simple dependence on the Q-value (see Eq. \ref{simpleest})
because of our assumption that the configuration mixing strength is 
nucleus-independent for $GT^+$ blocked nuclei.
 \label{highufec}}
\end{figure}

With the assumption that the high density electron capture rates
are dominated by transitions involving low-lying configuration 
mixing strength, these rates are
trivial functions of the electron Fermi energy and the 
parent-daughter mass difference. This is shown in figure \ref{highufec}
where the electron capture rates are presented 
for all nuclei with (N-Z)/A$>$0.1 that fall in the mass range 
A$=$66-80. The dependence of the rate on 
the Q-value for the transition is given by the simple analytic expression 
\begin{equation}
\label{simpleest}
\lambda \approx 2 \ln(2) 10^{-3.6} |M_{\rm GT}|^2({\rm F}_4(\eta_{\rm eff})+
\bar{q}^2{\rm F}_2(\eta_{\rm eff})+2\bar{q}{\rm F}_3(\eta_{\rm eff})).
\end{equation}
Here $\eta_{\rm eff}=(U_{\rm F}+m_e+q-E_{\rm res,d})/T$, with $E_{\rm res,d}$
the centroid of the ${\rm GT^+}$ resonance in the daughter, 
$\bar{q}=|q-E_{\rm res,d}|/T$, and the extra prefactor of 2 approximately
accounts for the Coulomb distortion of the incoming electron for nuclei
with $Z\approx 30-40$. This equation assumes that $q-E_{\rm res,d}$ 
is negative. The functions 
\begin{equation}
{\rm F_k(\eta)}=\int_0^{\infty} {x^kdx \over e^{x-\eta}+1}
\end{equation}
appearing in Eq. \ref{simpleest}
are the relativistic fermi integrals. Setting $E_{\rm res,d}=2{\, \rm MeV}$ and
$|M_{\rm GT}|^2=1$ in Eq. \ref{simpleest} gives a reasonable estimate
of the electron capture rates (when the rates are appreciable) when
$U_{\rm F}>10\,{\rm MeV}$. These expressions can be used in place of 
our rate tables under these conditions. Using the simple approximations to the fermi functions
developed in \cite{fuller4} gives an analytic approximation to the 
high density electron capture rates that is accurate to within the 
uncertainty in these rates. Electron capture rates for neutron rich
nuclei at high electron fermi energies are essentially a function of
only 1 parameter, the total $GT^+$ strength within a few MeV of the daughter 
state. Discrepancies between our calculated rates and more detailed calculations,
e.g. monte carlo+RPA calculations \citep{monte1}, reflect the 
difference between our adopted $|M_{\rm GT}|^2=1$ and the more 
detailed estimate of $M_{\rm GT}$.

Fig.  \ref{highufec} also gives an estimate of the uncertainties in
the electron capture rates. At $T_9=10$, $\rho Y_e=3\cdot 10^{10}{\rm g/cm^3}$
(${\rm U_f \approx 16 \,MeV}$), an error of 2 MeV in the position of the
centroid of the strength results in a change in the rate by a factor
of $\sim 4$ for the nuclei with the smallest rates, and a factor of
$\sim 2$ for the nuclei with the largest rates.  At $T_9=10$, $\rho
Y_e=10^{11}{\rm g/cm^3}$ (${\rm U_f \approx 23\,MeV}$), the uncertainty in the
placement of the centroid implies an error of at most a factor
of two in the electron capture rate. The uncertainty in the total
strength is probably about a factor of 2 on average.

\begin{figure}
\plotone{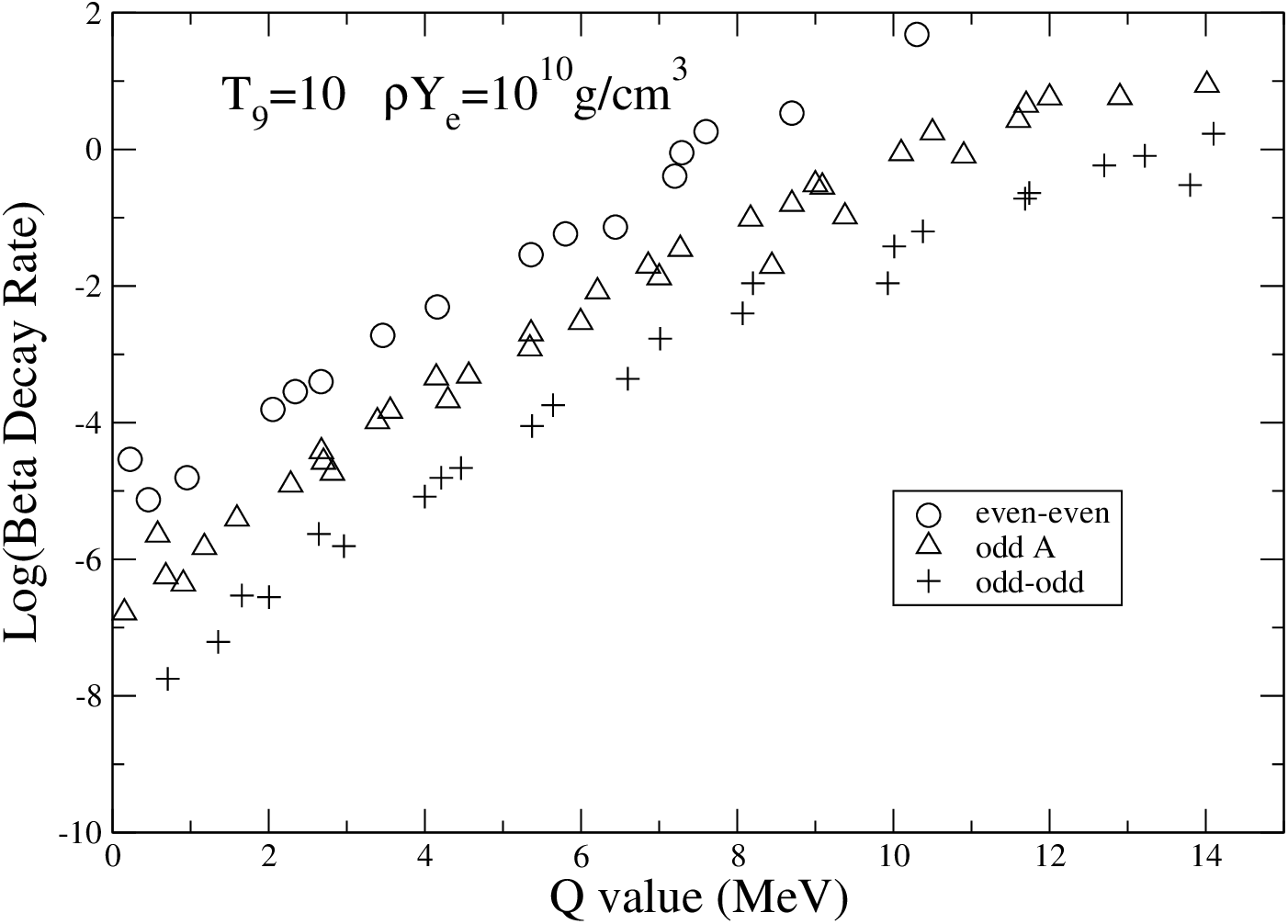}
\caption{$\beta$ decay rates for neutron rich nuclei in the mass range
A$<65\leq80$ at $T_9=10$ and $\rho Y_e={10^{10}}{\rm g/cm^3}$. The systematics
illustrated here are discussed in the text. \label{allb-.nrich}}
\end{figure}

The systematics of the $\beta$ decay rates at high temperature and
density are also simple. In Fig. \ref{allb-.nrich} we show the $\beta$
decay rates at $T_9=10$, $\rho Y_e=10^{10}{\rm g/cm^3}$ for all nuclei
with ${\rm (N-Z/A)>0.1}$. These rates fall into three distinct bands
corresponding to odd-odd nuclei, even-even nuclei, and
even-odd/odd-even nuclei. This can be understood by noting that at
$T_9=10$ the ${\rm GT^+}$ resonance is thermally populated (under our
assumption that the strength is centered at 2 MeV for these blocked
nuclei). In this case the rate of decay is approximately $\lambda({\rm
Q}) Z_{\rm daughter}/Z_{\rm parent}$, where $\lambda({\rm Q})$ has a
simple dependence on the Q-value. For even-even parents, $Z_{\rm
daughter}/Z_{\rm parent}$ is typically approximately 20 at $T_9=10$,
for odd-odd parents the ratio is about $1/20$, and for
odd-even/even-odd parents the ratio of partition functions is about 1.
Note that the $\beta$ decay rates are more sensitive to the (unknown)
details of the strength distribution than the electron capture rates.
At $T_9=10$, an error in the placement of the centroid of the
resonance of 2 MeV changes the beta decay rate by about an order of
magnitude.

\subsection{Discussion and Conclusions}

We have provided estimates of weak interaction rates for nuclei in 
the mass range A=65-80. These may be useful in simulations of x-ray bursts
and pre-supernova stellar evolution. The rates have been calculated 
using available experimental information and simple estimates for the
strength distributions and matrix elements for allowed and 
discrete-state transitions. The efficacy of our approach is confirmed 
through comparisons with detailed shell model based rates for 
nuclei in the mass range 60-65.

The single most uncertain aspect of the rate calculations is the ${\rm
GT^+}$ resonance strength distribution. Our simple prescription gives
good overall agreement with detailed shell model calculations of the
${\rm GT^+}$ strength for nuclei with A$\leq$65. However, our
prescription is probably not reliable for nuclei at the end of the fp
shell. Fortunately, nature to some extent does not seem to care about
some of the hard to get details of the strength distribution for these
nuclei. In the pre-collapse supernova this is because the nuclei
present are nearly blocked to $GT^+$ transitions. (n,p) exchange
experiments on such nuclei show that the strength is broad and low
lying in the daughter. Because the electron fermi energies are high in
the dense pre-supernova Fe-core, this implies that the electron
capture rate is principally a function of the
experimentally-determined parent-daughter mass difference. However,
our work does not provide the detailed estimates of the magnitude of
the configuration mixing strength that will ultimately also be needed for
supernova simulations.  The $\beta$ decay rates for the nuclei present
in the late-time pre-supernova core are exponentially sensitive to the
centroid energy of the ${\rm GT^+}$ resonance and are less certain. In
x-ray burst environments the opposite condition holds: the parent
daughter mass differences are typically large compared to the electron
energies, so that the decay rate is dominated by the
experimentally-determined lifetime.

The proper treatment of the high temperature electron capture and particulary
beta decay rates is a challenging and important issue. Special care is needed
in evaluating the partition functions for these rates. The beta decay 
rates for nuclei in these conditions is determined, among other things, 
by the nuclear partition 
function. In turn, the equilibrium nuclear composition is determined 
by the competition between beta decay and electron capture. If the
partition functions used to estimate the composition at a given $Y_e$ 
do not match with the partition functions used to calculate the weak 
rates, the calculated equilibrium of the sytem will not be correct.

\acknowledgments

The authors acknowledge helpful correspondence with G. Martinez-Pinedo
and K. Langanke regarding their calculations. We also thank Rob
Hoffman for useful discussions regarding the {\it rp}-process. This work was
partially supported by the DOE Program for Scientific Discovery
through Advanced Computing (SciDAC) at UCSD and LLNL and by NSF grant
PHY-00-0099499 at UCSD. A portion of this work was performed under the
auspices of the U.S. Department of Energy by University of California
Lawrence Livermore Laboratory under contract W-7405-ENG-48.

\clearpage

\begin{deluxetable}{crrrrrrrrrrr}
\tabletypesize{\scriptsize}
\tablecaption{Comparison of the shell model ${\rm GT^-}$ centroids calculated 
in Langanke (LMP) with those estimated from Eq. \ref{rpaminus} (Present). All
energies are in MeV and refer to energy with respect to the daughter
ground state.  
\label{tbl-1}}
\tablewidth{0pt}
\tablehead{
\colhead{Parent Nucleus}
& \colhead{LMP} 
& \colhead{Present}  
}
\startdata
$^{55}$Fe&12.6&10.4\\
$^{56}$Fe\tablenotemark{a}&9.6&8.5\\
$^{57}$Fe&12.6&11.5\\
$^{58}$Fe&11.0&9.5\\
$^{59}$Fe&13.6&12.6\\
$^{60}$Fe&10.3&10.6\\
$^{61}$Fe&13.8&13.75\\
$^{62}$Fe&11.8&11.7\\
$^{58}$Ni\tablenotemark{a}&9.2&6.44\\
$^{59}$Ni&10.6&9.46\\
$^{60}$Ni\tablenotemark{a}&9.0&7.5\\
$^{61}$Ni&13.3&10.6\\
$^{62}$Ni&9.2&8.2\\
$^{63}$Ni&13.2&11.5\\
$^{64}$Ni&9.6&9.15\\
$^{65}$Ni&12.&12.35\\
$^{56}$Co&13.2&12.7\\
$^{57}$Co&12.5&10.77\\
$^{58}$Co&14.7&13.7\\
$^{59}$Co&13.1&11.6\\
$^{60}$Co&14.0&14.8\\
$^{61}$Co&13.6&12.6\\
$^{62}$Co&15.3&13.9\\
$^{63}$Co&14.4&13.7\\
$^{64}$Co&16.0&16.7\\
$^{65}$Co&14.6&14.8\\
$^{55}$Mn&13.0&11.8\\
$^{56}$Mn&14.7&15.1\\
$^{57}$Mn&13.1&13.5\\
$^{58}$Mn&15.5&16.4\\
$^{59}$Mn&14.0&14.7\\
$^{60}$Mn&16.0&17.4\\
$^{61}$Mn&16.2&15.5\\
 \enddata


\tablenotetext{a}{For these nuclei the results from (p,n) experiments
give a centroid approximately 0.7-1 MeV lower than the shell model
results.}

\tablecomments{The shell model centroids in this table were taken from the
tables or estimated from the graphs in Caurier et al. 1999 and LMP.}

\end{deluxetable}

\clearpage

\begin{deluxetable}{crrrrrrrrrrr}
\tabletypesize{\scriptsize}
\tablecaption{Comparison of the shell model ${\rm GT^+}$ centroids calculated 
by LMP with those estimated from Eq. \ref{rpaplus} (Present). All
energies are in MeV and refer to energy with respect to the daughter
ground state.  
\label{tbl-2}}
\tablewidth{0pt}
\tablehead{
\colhead{Parent Nucleus}
& \colhead{LMP} 
& \colhead{Present}  
}
\startdata
$^{55}$Mn&4.6&4.6\\
$^{56}$Mn&5.9&6.4\\
$^{56}$Fe&2.6&2.4\\
$^{56}$Co&8.2&8.8\\
$^{58}$Mn&5.5&5.15\\
$^{58}$Co&7.35&8.1\\
$^{58}$Ni&3.75&3.65\\
$^{59}$Co&5.05&5.0\\
$^{60}$Co&6.35&6.74\\
$^{60}$Ni&3.4&2.7\\
$^{61}$Fe&2.1&1.8\\
$^{61}$Co&3.7&3.4\\
$^{61}$Ni&4.7&4.7\\
$^{61}$Cu&6.7&6.4\\
$^{62}$Ni&2.1&1.8\\
$^{64}$Ni&1.3&1.8\\
\enddata


\tablecomments{The shell model centroids in this table were also taken 
from Caurier et al. 1999 and LMP. Where an experimental result is listed along with
the shell model result 
we have presented the experimental result.}

\end{deluxetable}

\clearpage

\clearpage

\clearpage

\clearpage

\clearpage

\clearpage

\clearpage

\clearpage

\clearpage

\clearpage

\clearpage

\clearpage

\clearpage

\clearpage

\clearpage

\end{document}